\documentclass[3p,twocolumn,review,number,sort&compress]{elsarticle}

\usepackage{lineno,hyperref,amsmath,xcolor}
\usepackage{textcomp}
\graphicspath{{figures/}}
\modulolinenumbers[5]

\journal{Nuclear Instruments and Methods in Physics Research Section A}

\begin{document}
\newcommand{\mmcu}    {mm\ensuremath{^3}}
\newcommand{\cmsq}    {cm\ensuremath{^2}}
\newcommand{\degree}  {\ensuremath{^\circ}}
\newcommand{\msq}     {m\ensuremath{^2}}
\newcommand{\mum}     {\textmu m}
\newcommand{\myal}    {Al}
\newcommand{\myreg}   {\textsuperscript{\textregistered}}
\newcommand{\mymod}   {Module-0}
\newcommand{\myeta}   {\ensuremath{\eta}}
\newcommand{\myphi}   {\ensuremath{\phi}}
\newcommand{\mTPC}    {\ensuremath{\mu}TPC}
\newcommand{\myropan} {RO panel}
\newcommand{\myropans}{RO panels}
\newcommand{\myfig}[4]  {\begin{figure}[ht!] \centering
                        \includegraphics[width=#2mm]{#1}
                        \caption{#4.} \label{#1} \vspace{#3mm} \end{figure}
                        \typeout{<use Figure \thefigure>}}
\newcommand{\myfigd}[6] {\begin{figure}[ht!] \centering \includegraphics[width=#2mm]{#1}
                        \newline \vspace{#5mm} \includegraphics[width=#2mm]{#6}
                        \caption{#4.} \label{#1} \vspace{#3mm} \end{figure}
                        \typeout{<use Figure \thefigure>}}
\newcommand{\myfigt}[9]{\begin{figure}[ht!] \centering \includegraphics[width=#2mm]{#1}
                        \newline \vspace{#5mm} \includegraphics[width=#4mm]{#6}
                        \newline \vspace{#8mm} \includegraphics[width=72mm]{#9}
                        \caption{#3.} \label{#1} \vspace{2mm} \end{figure}
                        \typeout{<use Figure \thefigure>}}
\newcommand{\mybfig}[4] {\begin{figure*}[t!] \centering \includegraphics[width=#2mm]{#1}
                        \caption{#4.} \label{#1} \vspace{#3mm} \end{figure*}
                        \typeout{<use Figure \thefigure>}}
\newcommand{\figdum}[4] {\begin{figure}[ht!] \centering \includegraphics[width=#2mm]{dummy}
                         \caption{#4.} \label{#1} \vspace{#3mm} \end{figure}
                         \typeout{<use Figure \thefigure>}}
\newcommand{\figdeleted}{\begin{figure}[ht!] \centering \includegraphics[width=78mm]{dummy}
                         \caption{\myred{This figure has been deleted.}} \vspace{0.5mm} \end{figure}
                         \typeout{<use Figure \thefigure>}}
\newcommand{\myred}[1] {\textcolor{red}{#1}}
\newcommand{\mydummy}[1] {\relax}


\begin{frontmatter}

\title{Construction techniques and performances of a full-size prototype Micromegas chamber for the ATLAS muon spectrometer upgrade}


%
\author[ntua]{T.~Alexopoulos}
\author[napoli]{M.~Alviggi}
\author[frascati]{M.~Antonelli}
\author[roma1]{F.~Anulli}
\author[frascati]{C.~Arcangeletti}
\author[roma1]{P.~Bagnaia}
\author[roma3]{A.~Baroncelli}
\author[frascati]{M.~Beretta}
\author[roma1]{C.~Bini}
\author[cern]{J.~Bortfeldt}
\author[pavia]{D.~Calabr\`o}
\author[napoli]{V.~Canale}
\author[roma1]{F.~Capocasa}
\author[roma1]{G.~Capradossi}
\author[cosenza]{G.~Carducci}
\author[pavia]{A.~Caserio}
\author[napoli]{C.~Cassese}
\author[frascati]{S.~Cerioni}
\author[roma1]{G.~Ciapetti\fnref{fnguido}}
\author[roma3]{V.~D'Amico}
\author[napoli]{B.~De~Fazio}
\author[cosenza]{M.~Del~Gaudio}
\author[ntua,bnl]{P.~Gkountoumis}
\author[napoli]{C.~Di~Donato}
\author[frascati]{R.~Di~Nardo}
\author[roma1]{D.~D'Uffizi}
\author[pavia]{E.~Farina}
\author[pavia]{R.~Ferrari}
\author[pavia]{A.~Freddi}
\author[frascati]{C.~Gatti}
\author[pavia]{G.~Gaudio}
\author[lecce]{E.~Gorini}
\author[lecce]{F.~Gravili}
\author[pavia]{S.~Guelfo~Gigli}
\author[ntua]{G.~Iakovidis} 
\author[cern]{P.~Iengo}
\author[lecce]{A.~Innocente}
\author[pavia]{G.~Introzzi}
\author[roma3]{M.~Iodice}
\author[ntua,bnl]{E.~Karentzos} 
\author[ntua,bnl]{A.~Koulouris}
\author[pavia]{A.~Kourkoumeli-Charalampidi \corref{mycorrespondingauthor}}
\author[roma1]{F.~Lacava}
\author[pavia]{A.~Lanza}
\author[cosenza]{L.~La~Rotonda}
\author[frascati]{S.~Lauciani}
\author[roma1]{L.~Luminari}
\author[frascati]{G.~Maccarrone}
\author[ntua]{S.~Maltezos} 
\author[frascati]{G.~Mancini \corref{mycorrespondingauthor}}
\author[roma3]{L.~Martinelli}
\author[napoli]{P.~Massarotti}
\author[lecce]{A.~Miccoli}
\author[lecce]{A.~Mirto}
\author[ntua,bnl]{P.~Moschovakos}
\author[ntua,bnl]{K.~Ntekas}
\author[cosenza]{S.~Palazzo}
\author[roma3]{G.~Paruzza}
\author[roma3]{F.~Petrucci}
\author[pavia]{L.~Pezzotti}
\author[frascati]{G.~Pileggi}
\author[roma1]{A.~Policicchio}
\author[napoli]{G.~Pontoriere}
\author[frascati]{B.~Ponzio}
\author[pavia]{E.~Romano}
\author[cosenza]{V.~Romano}
\author[napoli]{L.~Roscilli}
\author[pavia]{G.~Rovelli}
\author[pavia]{C.~Scagliotti}
\author[cosenza]{M.~Schioppa}
\author[napoli]{G.~Sekhniaidze}
\author[roma3]{M.~Sessa}
\author[pavia]{S.~Sottocornola}
\author[cosenza]{P.~Turco}
\author[roma1]{M.~Vanadia}
\author[roma1]{D.~Vannicola}
\author[frascati]{T.~Vassileva}
\author[roma3]{V.~Vecchio}
\author[pavia]{F.~Vercellati}
\author[roma1]{A.~Zullo}
\cortext[mycorrespondingauthor]{Corresponding author}
\address[ntua]{National Technical University of Athens, Heroon Polytechniou 9, Zografou, Greece.}
\address[napoli]{Dipartimento di Fisica, Universit\`a di Napoli, Napoli e INFN Sezione di Napoli, Napoli; Italy.}
\address[frascati]{INFN e Laboratori Nazionali di Frascati, Frascati (Roma); Italy.}
\address[roma1]{Dipartimento di Fisica, Sapienza Universit\`a di Roma, Roma e INFN Sezione di Roma, Roma; Italy.}
\address[roma3]{Dipartimento di Matematica e Fisica, Universit\`a Roma Tre, Roma e INFN Sezione di Roma Tre, Roma; Italy.}
\address[cern]{CERN, Geneva, Switzerland.}
\address[pavia]{Dipartimento di Fisica, Universit\`a di Pavia, Pavia e INFN Sezione di Pavia, Pavia; Italy.}
\address[cosenza]{Dipartimento di Fisica, Universit\`a della Calabria, Rende e INFN Gruppo Collegato di Cosenza, Laboratori Nazionali di Frascati; Italy.}
\address[lecce]{Dipartimento di Matematica e Fisica, Universit\`a del Salento, Lecce e INFN Sezione di Lecce, Lecce; Italy.}
\address[bnl]{Brookhaven National Laboratory, Upton, NY 11973-5000, U.S.A.}
\fntext[fnguido]{The work described here was one of Guido's latest contributions to ATLAS. The authors wish to dedicate this article to him, in memory of all his activities.}

\begin{abstract}

%

A full-size prototype of a Micromegas precision tracking chamber for the upgrade of the ATLAS detector at the LHC Collider has been built between October 2015 and April 2016.
This paper describes in detail the procedures used in constructing the single modules of the chamber in various INFN laboratories and the final assembly at the Laboratori Nazionali di Frascati (LNF).
Results of the chamber exposure to the CERN SPS/H8 beam line in June 2016 are also presented.
The performances achieved in the construction and the results of the test beam are compared with the requirements, which are imposed by the severe environment during the data-taking of the LHC foreseen for the next years.

\end{abstract}


\end{frontmatter}

\nolinenumbers

\section{Introduction} 
\label{intro}

\par In order to benefit from the expected high luminosity that will be provided by the Phase-I and High-Luminosity upgrades of LHC \cite{uplhc}, the innermost station of the ATLAS Muon Spectrometer in the forward region (Small Wheel) will be replaced.
The new detectors will operate in a high background radiation environment (up to 15 kHz/\cmsq) while reconstructing muon tracks with high precision as well as providing information for the Level-1 trigger.
These performance criteria are demanding.
In particular, the precision reconstruction of tracks for offline analysis requires a spatial resolution of about 100 \mum\ per detector layer, and the Level-1 trigger track segments have to be reconstructed online with an angular resolution of approximately 1 mrad.
For this purpose a new set of detectors have been designed and are being constructed for the New Small Wheel (NSW) \cite{nswtdr}.
The NSW chambers use two technologies, one primarily devoted to the Level-1 trigger function (small-strip Thin Gap Chambers, sTGC \cite{stgc}) and one dedicated to precision tracking (Micromegas detectors, \emph{MM} \cite{microm}).
The MM detectors have outstanding precision tracking capabilities due to their small gap (5 mm) and strip pitch (approximately 0.5 mm).
\mydummy{Such a precision is crucial to maintain the current ATLAS muon momentum resolution in the high background environment of the upgraded LHC running conditions.}
The combined information of the MM and sTGC chambers can also be used in the Level-1 trigger to confirm the existence of track segments found by the muon end-cap middle station (Big Wheels).

\mybfig{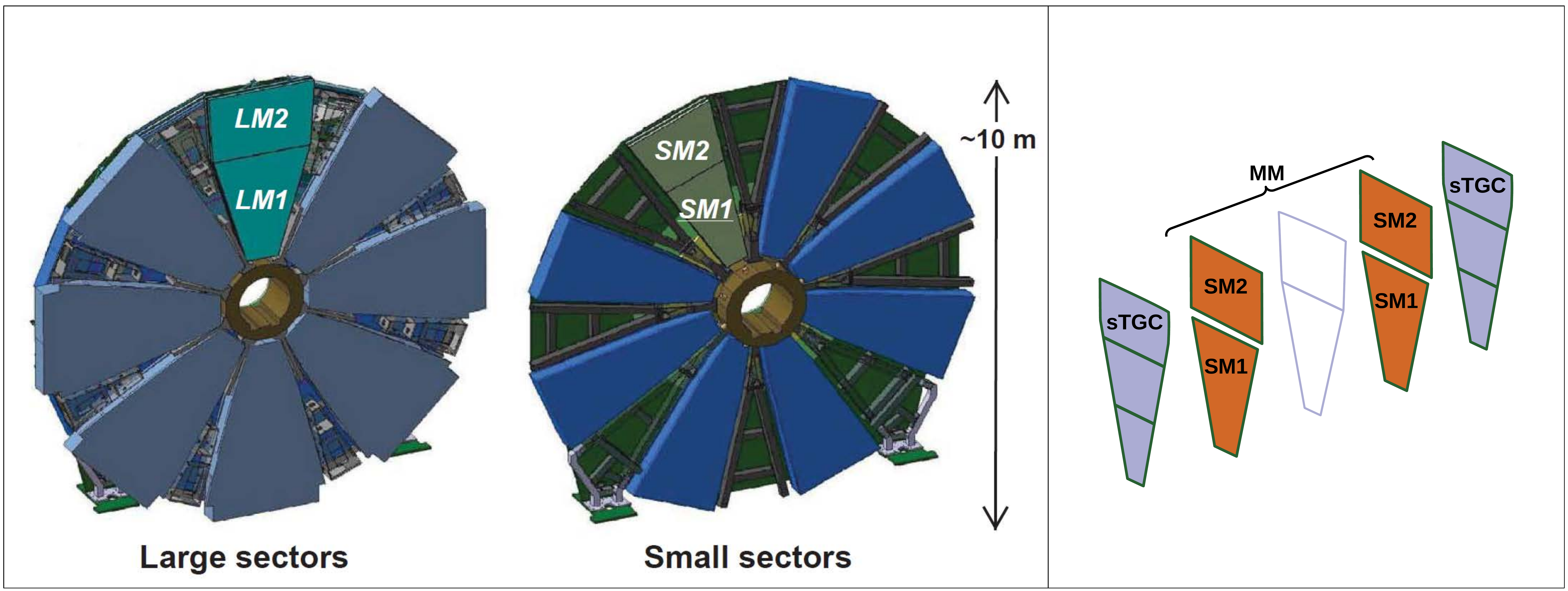}{160}{0}{Overall view of the New Small Wheel.
Left: views of the wheel, highlighting the large sectors LM1 and LM2 and the small ones SM1 and SM2.
Right: exploded view of a SM1 module, with two sTGC chambers and two MM quadruplets}

\par The realization of the MM is shared between four groups of laboratories, one for each type of chambers (see Sec.\ref{layout}).
In this paper the construction and the performance of the first full-size MM chamber is presented.
This detector, referred in the following as \emph{Module-0}, was built by the INFN consortium as a full size prototype of the chambers located in the small-sectors at large pseudorapidity \myeta\ \footnote{We use here the LHC notation $\eta = - \ln[\tan(\theta/2)]$ with $\theta$ the polar angle of the track originated from the interaction point.} (called SM1, see Sec.\ref{layout}).
The construction started in late October 2015 and was completed in April 2016.
The performances were checked in a test beam at CERN in June 2016.

\par In the following we present at first the general layout of the chamber (Sec.\ref{layout}).
Then the construction procedures for the different parts of the chamber, the read-out panels (Sec.\ref{ro-panel}), the drift panels (Sec.\ref{drift-panel}), the mesh including its gluing on the drift panels (Sec.\ref{drift-comple}), and the chamber assembly (Sec.\ref{assembly}) are described.
The performance, measured at a test beam facility, are described in Sec.\ref{testbeam}.
Finally the conclusions are shown in Sec.\ref{conclu}.

\section{Module Layout} 
\label{layout}

\par The overall structure of the New Small Wheel \cite{nswtdr} is shown in Fig.\ref{layout_1}.

\subsection{The New Small Wheel} 

\par Each of the two wheels consists of eight large and eight small sectors partially overlapping, housing sTGC and MM chambers.
For what concerns the MM, each sector is segmented in two parts of different size, covering different regions in \myeta.
This division results in four kind of chambers: SM1 (the subject of this paper) and SM2 as Small Sectors Modules, LM1 and LM2 as the Large Sectors Modules.
The chamber size is $\sim$2 \msq (SM1 and SM2) and $\sim$3 \msq (LM1 and LM2).

\mybfig{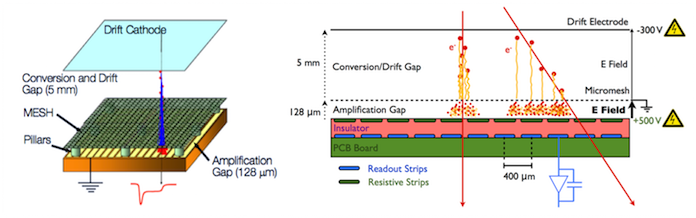}{150}{0}{A scheme of a single MM layer.
Left: the exploded view of a single gap unit.
Right: the signal production and read-out}

\par The MM tracking system is designed with eight independent layers, orthogonal to the LHC beam-line, organized in two \emph{quadruplets}, as shown in Fig.\ref{layout_1}.
Muon tracks coming from the LHC interaction point are expected to impinge on each layer with angles in the range $8\degree\div32\degree$ with respect to the direction orthogonal to the layers.
This design provides eight precision points per muon track, with an overall lever arm of about 20 cm.
As already stressed, for each point a spatial resolution of $\sim$100 \mum\ is required.

\par The construction of the 32 SM1 chambers (16 for each wheel) has been assigned to the INFN sections of Cosenza, Frascati, Lecce, Napoli, Pavia, Roma Tre, Roma Sapienza.

\subsection{The Micromegas chamber} 

\par A single MM layer is shown in Fig.\ref{layout_2}.
Two Printed Circuit Boards (\emph{PCB}) create a uniform electric field ($\sim$600 V/cm) in a 5 mm gas gap.
The drift cathode (the \emph{drift PCB}) is held at negative HV (typically $-300$ V) with respect to a stainless steel mesh, which is grounded.
The mesh is stretched above a series of 128 \mum\ high pillars posed on the anode, composed of resistive strips on a kapton\myreg ~layer held at positive HV (typically around 500-600 V).
The resulting electric field between the mesh and the anode is very large ($\sim$50 kV/cm), more than 50 times higher than the drift field.
This scheme guarantees the almost complete transparency of the mesh and the almost complete evacuation of the avalanche ions in a short time ($\sim$100 ns) in the mesh.
Below the kapton, metallic strips capacitively coupled with the resistive strips allow to read out the electric signals \cite{res}.
Pillars, resistive and metallic read-out strips are integrated in the \emph{read-out PCB}s.

\subsection{The Quadruplet} 

\myfig{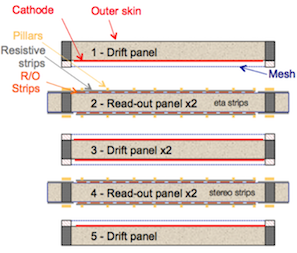}{77}{5}{A schematic view of the five panels of a MM quadruplet}

\myfig{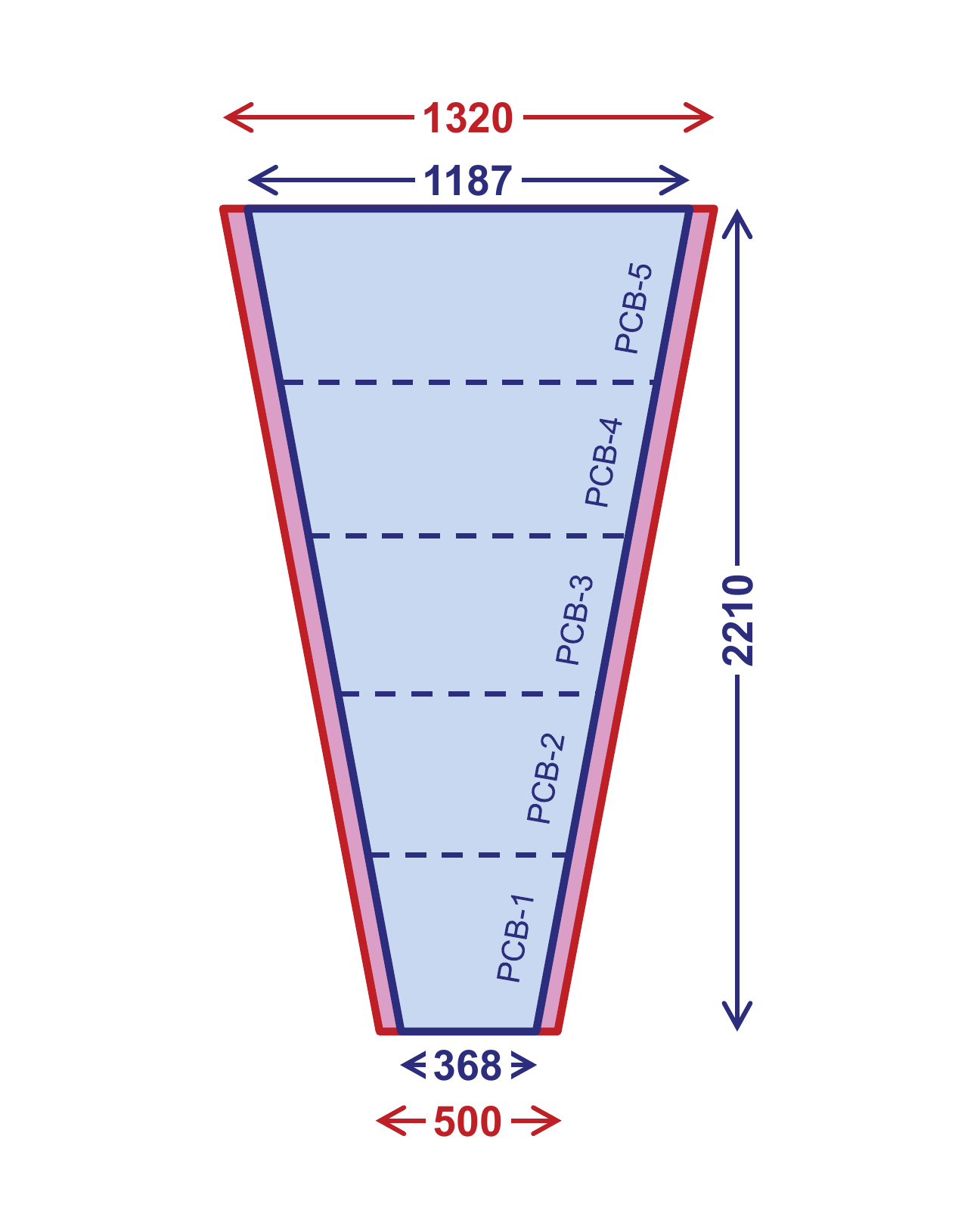}{77}{5}{The shape of a SM1 quadruplet. In blue the drift panels and in red the extensions of the read-out panels. Dimensions are in mm}

\par Fig.\ref{layout_3} shows a schematic view of a quadruplet.
The four active gaps including the meshes are bounded by five panels, providing the required stiffness.
The panels are trapezoidal in shape, 11.77 mm thick, 2210 mm high, 368 mm and 1187 mm wide at the two extremities (Fig.\ref{layout_4}).
As also shown in the figure, the read-out panels have an extra border for the front-end electronic boards and services, such that their dimensions extend to 500 mm and 1320 mm at the two extremities.
All panels consist of two sets of PCBs, facing an aluminum (\myal) honeycomb layer 10.1 mm thick with 6 mm hexagonal cells.
For construction purposes a panel is produced by connecting together smaller PCBs (PCB1-5, Fig.\ref{layout_4}).
The structure is given by \myal\ bars with 10.0 mm thickness, both as perimetric frames and as reinforcement beams.
The PCBs are glued\footnote{Araldite 2011\myreg\ was used for both the panels and the mesh.} onto the honeycomb and the bars.

\par Three out of the five panels (panels 1, 3 and 5 in Fig.\ref{layout_3}), which contain drift cathodes only, are the \emph{drift panels}.
For the drift panel production only the inner FR4 face of the external drift panels (1 and 5 in figure 3) and both FR4 faces in the central drift panel (3 in figure) are covered with a 17 \mum\ copper layer providing the cathode for each of the four layers. In the present prototype also the outer face of the two external drift panels were covered with copper.

\par The other two panels (2 and 4) are the \emph{read-out panels}.
The Readout boards, the most important and critical part of the detector, are constituted by an FR4 layer with etched readout copper strips, resistive strips and pillars, giving rise to the read-out PCBs.
In the quadruplet layout, the strips of the first two layers (panel 2 in Fig.\ref{layout_3}, the \emph{eta panel}) are parallel to the chamber bases, and almost orthogonal to the bending plane of the tracks in the ATLAS Experiment.
For the third and fourth layer (panel 4, the \emph{stereo panel}) the strips are inclined by $\pm 1.5 \degree$\ with respect to the strips of the \emph{eta panel}.
This configuration allows for a precise determination of the coordinate \myeta, necessary for the measurement of momentum, still retaining a less precise determination of the second coordinate \myphi.

\par More specifically, the active area of the read-out PCB consists of a 500 \mum\ thick FR4 layer on top of which the Cu read-out strips are printed.
The Cu strips are 17 \mum\ thick and 300 \mum\ wide, while the strip pitch is 425 \mum.
Over the read-out strips a layer of Akaflex glue 25 \mum\ thick is deposited with a kapton foil 50 \mum\ thick over it, in order to isolate the resistive from the Cu strips.
The resistive strips are 300 \mum\ wide while the strip pitch is 425 \mum.
The final layer over the resistive strips consists of pyralux\myreg\ coverlay pillars 128 \mum\ high, maintaining a constant distance between the mesh and the resistive strips where the signal is amplified \cite{microm}  \cite{kobe}.
The pillars are placed at a distance of 7 mm between each other while their diameter is 300 $\mu$m.
The pillar shape and spacing was optimized to reduce the dead area while maintaining a mesh planarity, with a sufficient area at the bottom of the pillar to guarantee a successful adhesion to the PCB.

\par Both the drift and read-out PCBs are manufactured by the industry with the resistive layers being produced by Kobe University in Japan \cite{resist} and delivered to the PCB contractor.
\footnote{\label{noteltos}For the SM1 modules both the read-out and drift PCB were provided by \emph{Eltos S.p.A., Strada E, 44 - San Zeno - 52100 Arezzo (Italy)}.}.

\subsection{Quality requirements} 
\label{toler}

\par The quality of the MM chambers critically depends on their accurate construction.
The alignment of the strips of the read-out panels affects the precision in the measured coordinate.
The requirements in X (along the strips) and Y (on the strip plane, perpendicular to them) are:
\begin{enumerate}[a)]
\item 20 \mum\ in Y for the strip alignment of a single PCB;
\item 35 \mum\ in Y for the alignment between different PCBs;
\item 300 \mum\ for the alignment in X;
\end{enumerate}
\par \noindent Moreover, the uniformity of the electric field requires a good planarity of the panel and the correct grounding.
The nominal tolerances used for the panel planarity are:
\begin{enumerate}[a)] \setcounter{enumi}{3}
\item 37 \mum\ in RMS, equivalent to $\pm$ 110 \mum\ mechanical tolerance, for both read-out and drift panels;
\item 50 \mum\ for the bars and for the honeycomb.
\mydummy{\item 500 \mum\ on the outer faces for drift PCBs.}
\end{enumerate}

\par \noindent In addition, the electrical connection between the lateral frames, the inner bars and the honeycomb was carefully checked.

\section{Assembly and test of the Read-Out Panels} 
\label{ro-panel}

\par The two read-out panels (i.e. panels 2 and 4 in Fig.\ref{layout_3}) were prepared in a class 10,000 clean room of the \emph{Sezione INFN di Pavia} using a \emph{stiff-back technique}, developed in the workshop and extensively tested on several small panel prototypes.

\subsection{Construction and Measurement System} 

\mybfig{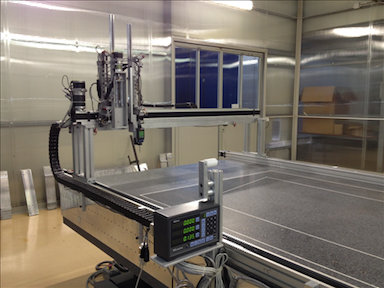}{130}{0}{The granite table and its support bridge}

\par The clean room was equipped with a (3500 $\times$ 2000 $\times$ 350) \mmcu\ granite table (with a certified maximum deviation of 8 \mum), on which a measurement system and the construction tools were placed.
The CMM (Coordinate Measuring Machine), a device for dimensional measurements which can reach an accuracy of 3 \mum\ over the entire table surface, was used for this purpose.
It is composed of two stainless-steel rails positioned along the longest sides (X axis), moving over the granite table a support bridge (Fig.\ref{ro_panel_1}), which holds a third rail (Y axis).
This rail controls the position of an arm, on which a height gauge and a glue dispenser are installed (Fig.\ref{ro_panel_2}).
The arm has an additional degree of freedom (Z axis) and the movements along the three axes are provided by remotely-controlled step motors.
The height gauge is composed by an indicator and an optical line\footnote{Mitutoyo Digimatic ID-H 563-561D\myreg\ and Mitutoyo AT-116 539-276-30\myreg.}.
The glue distribution is automated by using a remotely controlled dispenser with a cartridge, operated by an additional step motor.

\myfig{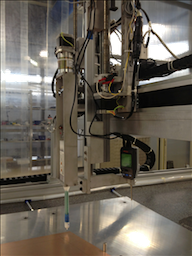}{76}{0}{The height gauge of the right and the glue dispenser on the left}

\par The construction of the panels is based on reference plates located on the granite table and the stiff-back.
A set of 5 milled \myal\ plates, (1450 $\times$ 490 $\times$ 20) \mmcu\ each, is placed on the granite table parallel to its short side and used as reference plane.
Due to the practical impossibility of machining a single surface of approximately 3.5 \msq\ within the required planarity (20 \mum), we opted for using 5 separate PCBs, blocked and aligned by means of reference pins positioned with precision holes in the granite table.
In order to allow for vacuum sucking with a pump, a set of pass-through holes and grooves were machined at the bottom side of all the plates.
Vacuum sucking is used to block the 5 PCBs (one on each reference plate) composing the lower surface of each panel in position on the reference plates for gluing.

\par The stiff-back is composed of 5 similar reference plates, mounted on a support made of two \myal\ skins spaced by 10 cm \myal\ honeycomb.
The stiff-back is attached to a crane (Fig.\ref{ro_panel_3}) and can be moved horizontally, vertically, and rotate to allow for an upward or downward facing of the RO plane on it, depending on the assembly phase.

\myfig{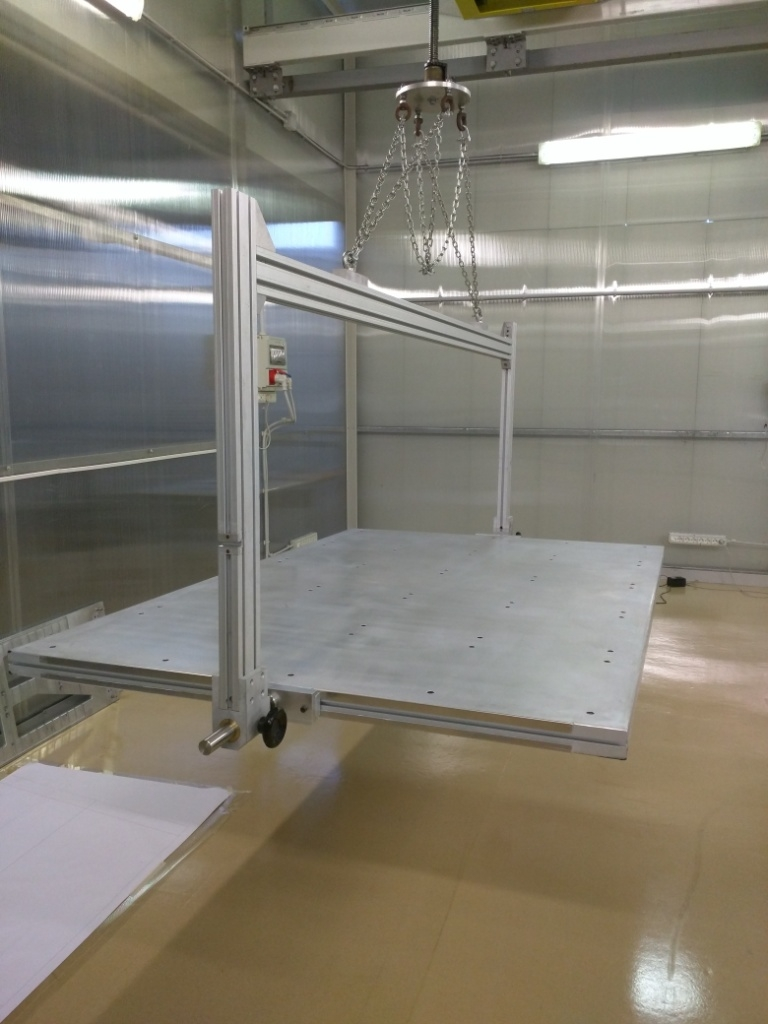}{76}{0}{The stiff-back}

\par Since the 10 plates define the two reference surfaces for construction and measurements of the components, a flatness check on them was performed.
As a result of 5 different scans, each consisting of 325 Z measurements on the reference surface, an average Z distribution and a repeatability map were obtained, both for the plates on the granite table and for the plates on the stiff-back. Each Z value is calculated as difference of the corresponding Z measurements (reference plate on the granite table minus granite table), to cancel out system distortions.

\subsection{Component QA/QC} 

\par A careful dimensional QA/QC on each component was performed before construction, in order to detect faulty components and to verify that all the mechanical specifications had been met.

\par The \myal\ frames positioned in the internal structure of the panels were checked both for their dimensions, especially thickness, and for possible deviation from straightness (bending and torsion).
The quality control aims at discarding frames outside tolerance, which may affect the planarity of the panel if used for construction.
The frames were measured on a granite table with a linear height\footnote{Mitutoyo LH-600E\myreg.} on all the sides, at a distance of 10 cm along their length.
The real dimensions and deformations were then checked.

\par The honeycomb sheets were only checked for thickness with a micrometer.
Due to the construction method described in the following section, a very precise height measurement of the components is not needed, since the discrepancies are compensated by layers of glue.
Still, there should be no interference with the PCBs (honeycomb too thick) or a too small distance between the PCBs themselves (honeycomb too thin).

\par A condition to ensure a good parallelism of the panels is the thickness uniformity of each PCB.
Therefore, the 20 PCBs received from the manufacturer were individually checked. In order to build a panel, 5 PCBs have been placed on the reference granite plate with an under pressure of 150 mbar, and their thickness has been measured using the CMM machine. 
The same measuring procedure was followed for the 5 PCBs positioned on the stiff-back plates.

\subsection{Panel Construction} 

\par As first step for the panel construction, the PCBs (PCB1-5 of Fig.\ref{layout_4}) are placed on the reference plates on the granite table and five equivalent PCBs on those of the stiff-back, precisely positioned using reference pins, and blocked by turning the vacuum on.
Internal and external frames, honeycomb and cooling bars are placed on top of the five PCBs on the granite table (Fig.\ref{ro_panel_4}).
A {\em dry-run} (full closure operation but the glue deposition) is performed to check the coupling of the stiff-back with the table reference plates: the stiff-back is rotated upside down (PCBs facing down) and moved over the table, then lowered on top of the reference plates.

\myfig{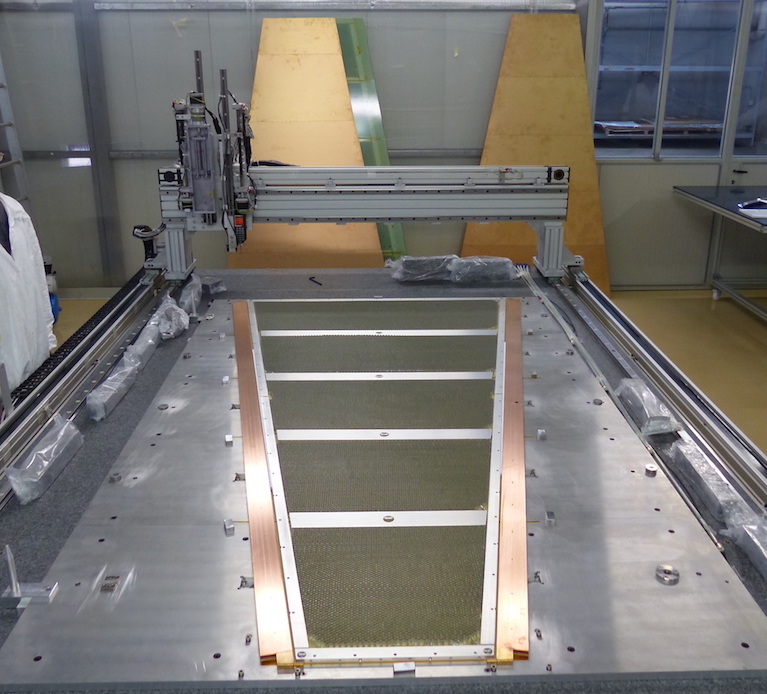}{76}{0}{A \myropan\ during construction}

\par The plates on the table and on the stiff-back are precisely matched by using a tapered interlock and a V-shaped interlock, which allow the correct positioning of the two PCB layers in a XY plane. In addition, 18 steel flat supports are used to guarantee the correct Z distance between the reference plates of the stiff-back and those of the table. After contact between the supports, reference measurements are taken
along the perimeter of the gap between the table reference plates and the stiff-back plates.

\par The stiff-back is then removed, and turned upside down once again (PCBs facing up).
At this point, the gluing of the two RO planes, frames, honeycomb and cooling bars is performed.
The glue is automatically disposed by a motor-controlled machine both on the PCBs on the table and those on the stiff-back.
The panel is closed by rotating the stiff-back upside down (PCBs facing down), moving it over the table, and finally lowering it on top of the reference plates.
The system is left untouched for the rest of the day (18 hours), for glue curing.

\par The following day, the stiff-back vacuum is turned off and the stiff-back removed from the table, leaving the glued panel still sucked on the reference plates of the table (Fig.\ref{ro_panel_5}).
The planarity of the panel is measured by means of the CMM.
Before switching off the table vacuum, alignment inserts (for \myeta-panels) or pins (for stereo panels) are glued to the panel.
The planarity of the panel is then measured again with vacuum off, to verify possible distortions due to internal tension of the panel.

\myfig{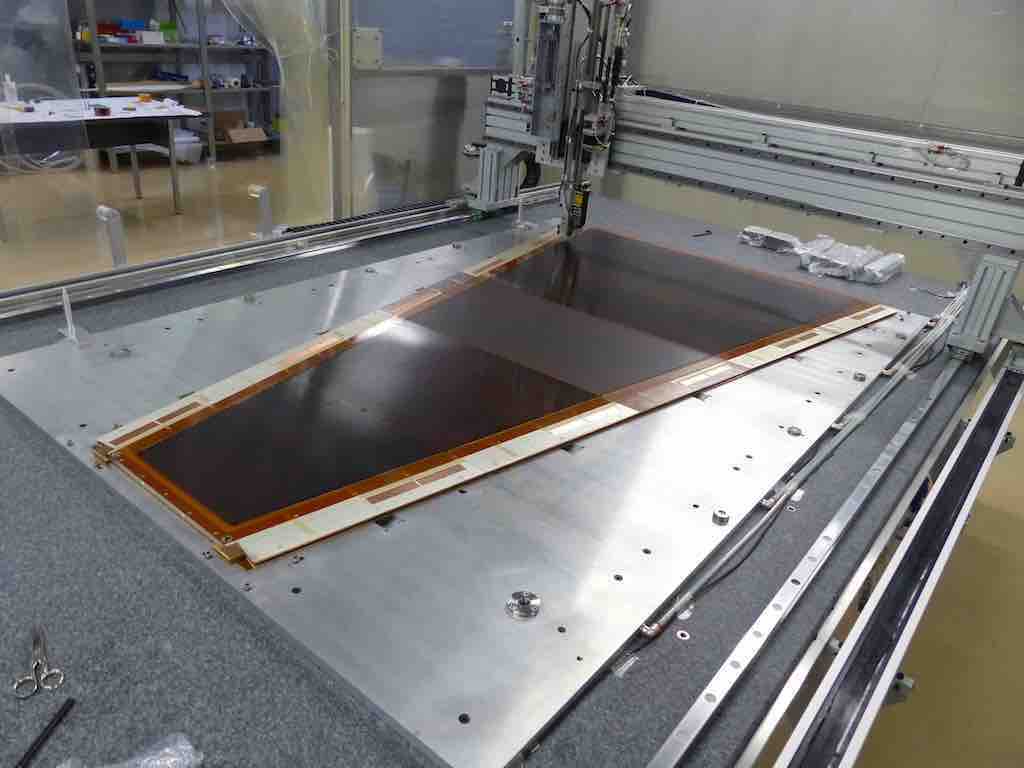}{76}{0}{A \myropan\ after the construction}

\subsection{\myropan\ QA/QC} 

\par The planarity of the two panels was measured by means of the CMM.
The thickness (Z$_{\rm mean}$ = 11.77 mm) and planarity (RMS = 24 \mum) of the stereo panel are shown in Fig.\ref{ro_panel_6}.
The measured RMS is significantly smaller than the required value of Sec.\ref{toler}, allowed by the project mechanical specifications.
The deviations from Z$_{\rm mean}$ measured on the stereo panel are shown in Fig.\ref{ro_panel_7}.
The range is (Z$_{\rm mean}$ - 70 \mum) $\leq$ Z $\leq$ (Z$_{\rm mean}$ + 55 \mum), well within the allowed tolerance.

\myfig{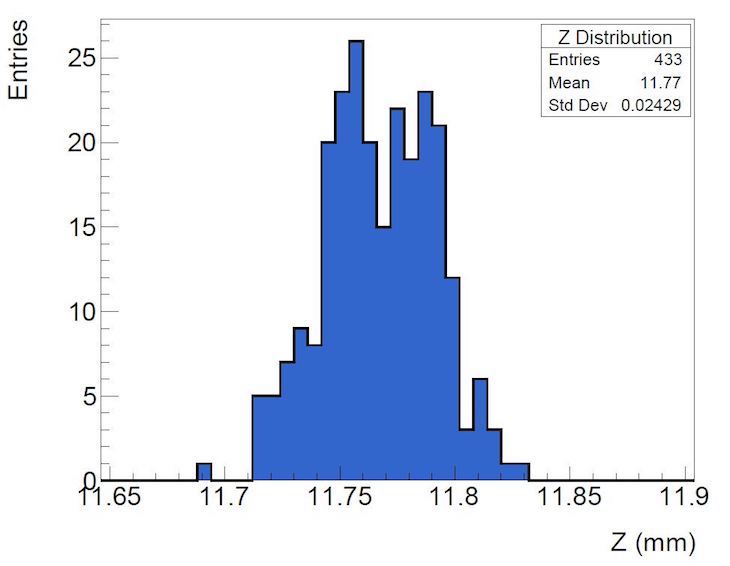}{76}{0}{Thickness and planarity of the stereo panel}

\myfig{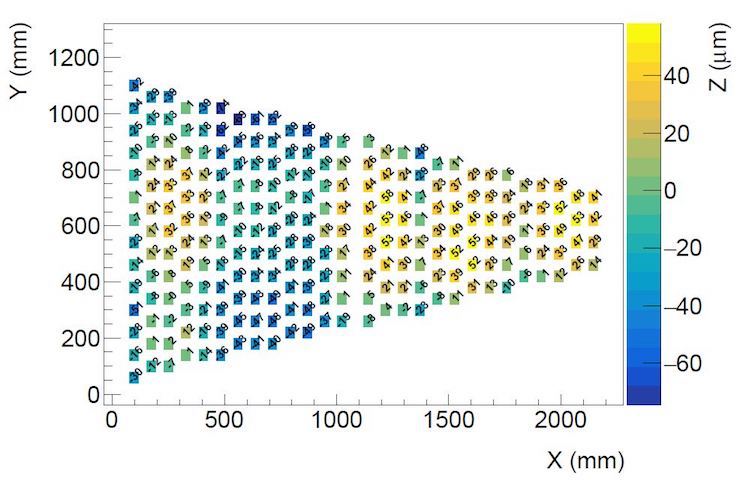}{76}{0}{Deviations from planarity for the stereo panel}

\par A set of additional tests will be performed on \myropans\ to be used in MM chambers for the NSWs: a strip alignment test, a test to check for possible gas leakage, and some electrical tests will be carried out.

\section{Assembly and test of the Drift Panels} 
\label{drift-panel}

\par The three drift panels of the \mymod\ (i.e. panels 1, 3, 5 in Fig.\ref{layout_3} were prepared in a class 10,000 clean room of the \emph{Sezione INFN di Roma} using the so called \emph{vacuum bag technique} \cite{vacuumbag} (Fig.\ref{drift_panel_1}), developed in the workshop and extensively tested in the preparation of several small panel prototypes.

\mybfig{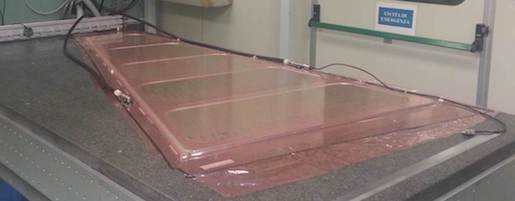}{140}{0}{A drift panel during glue curing (vacuum bag technique) on the granite table}

\subsection{Preliminary tests} 

\myfigt{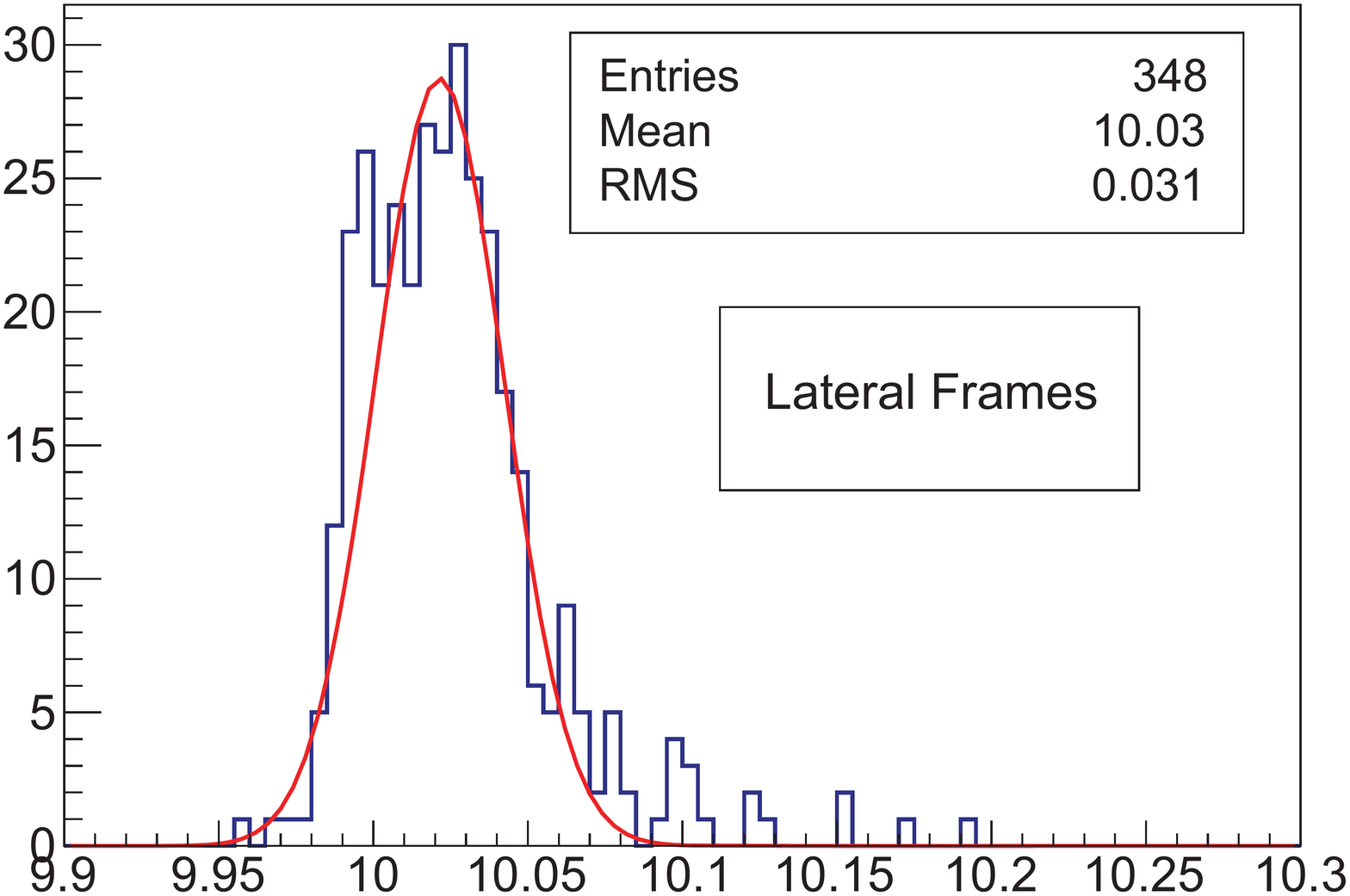}{77}{Measurements of the frames, PCBs and honeycomb of a drift panel (see text)}{73}{2}{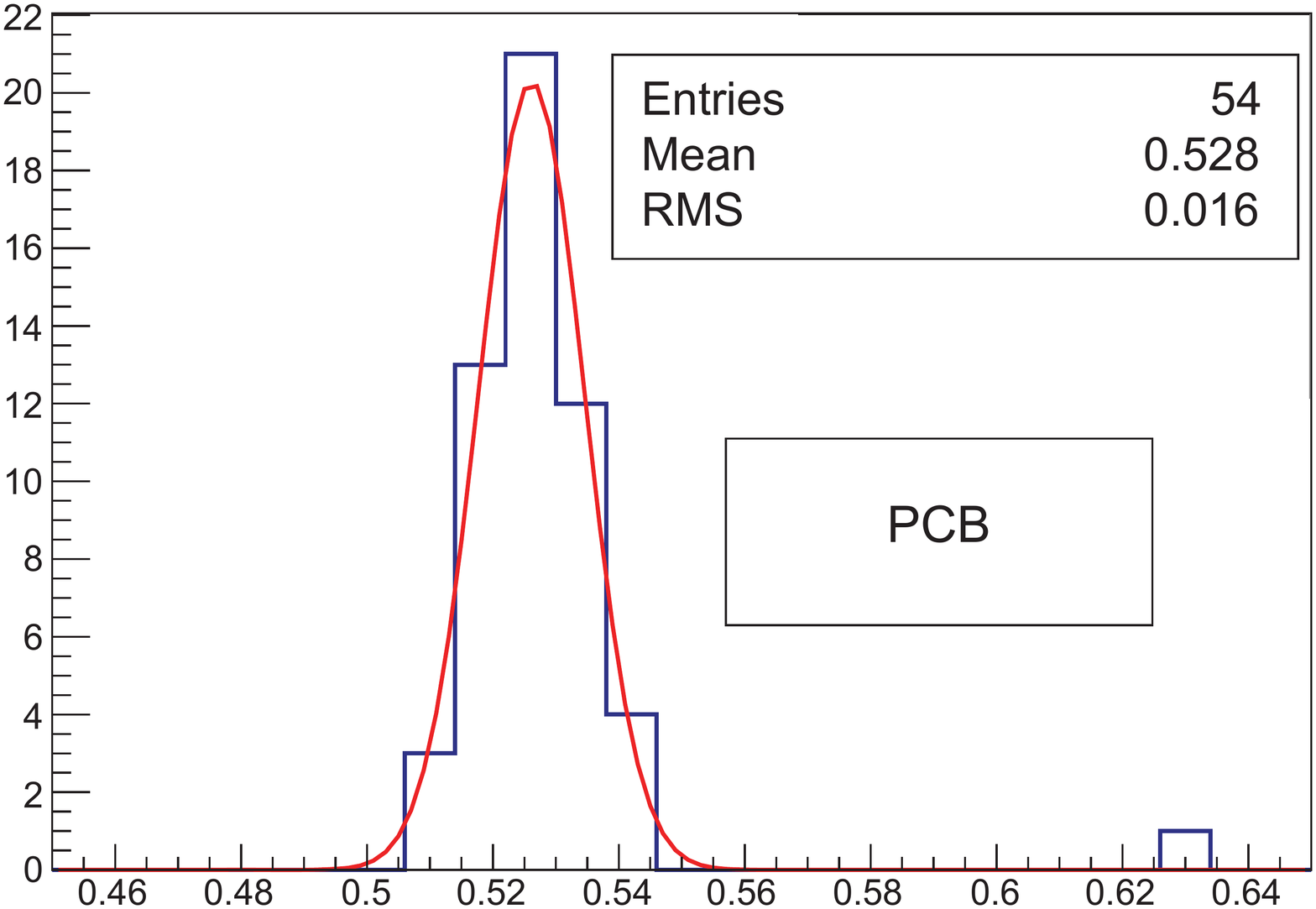}{72}{0.5}{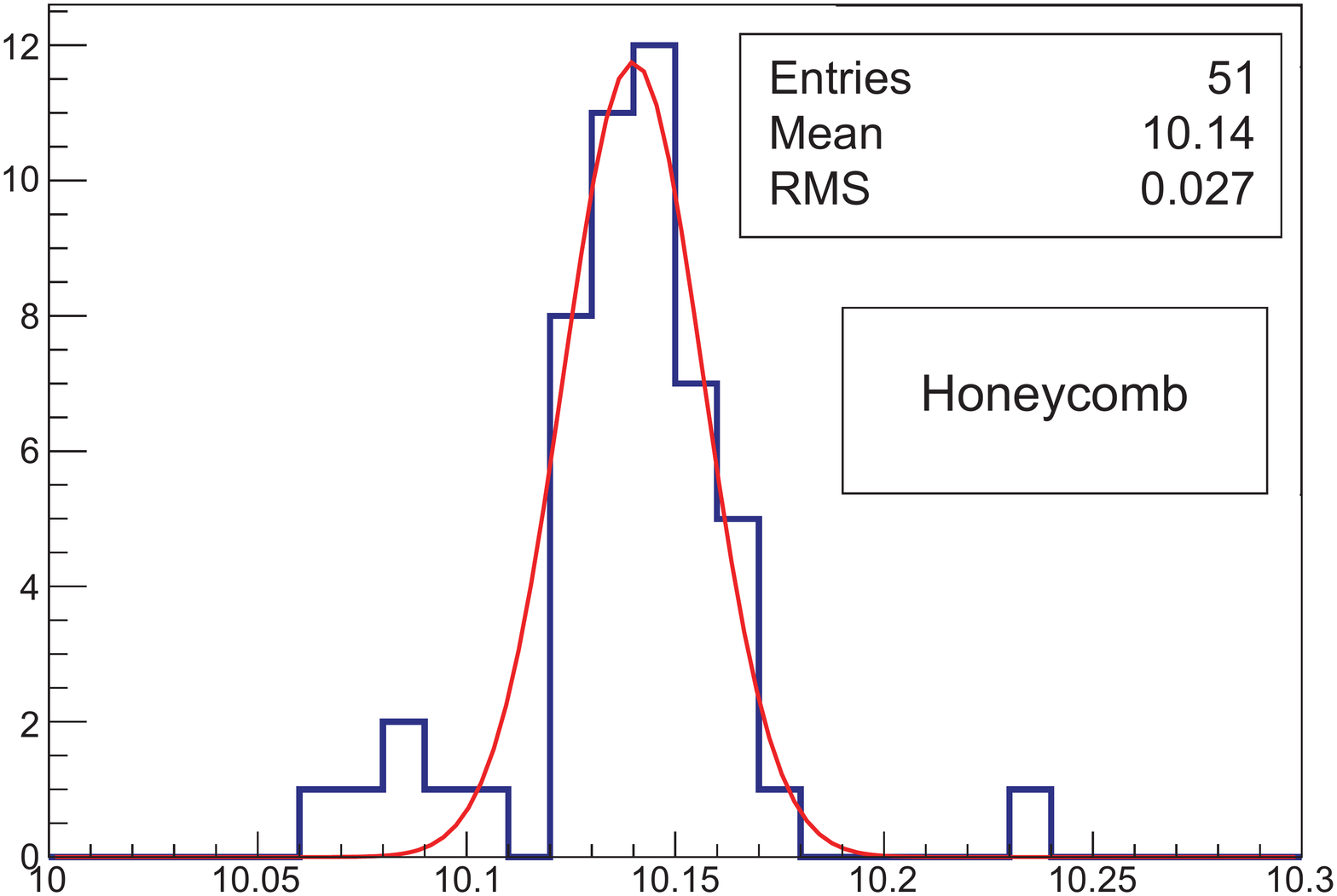}

\par The structure of the drift panels is described in Sec.\ref{layout}.
When using the vacuum bag technique the height and the planarity of the panel are determined by the pile-up of the glued components.
It is therefore mandatory to check the thickness of all the components before assembly.
The measurements were performed using the limbo tool, described in Sec.\ref{drift-tests}.
The distributions of the measurements, regularly spaced over their full length and surface, are given in Fig.\ref{drift_panel_2}.
For a drift panel, the upper plot shows the data for the lateral frames, the medium one the data for the PCBs, and the lower one the data for the honeycomb pieces, all measured along their edge.

\subsection{Drift panel assembly} 

\par The vacuum bag method provides a simple technique for panel assembly that takes advantage of the good planarity (certified up to a few microns) of the granite table in the clean room of the workshop.
Two variants of this method were used: the \emph{single-step} procedure for the outer drift panels and the \emph{two-steps} for the central drift panel.

\par The day before the assembly of a panel all the components were cleaned by alcohol and a dry-run, similar to the one already discussed in Sec.\ref{ro-panel}, was performed to check the components.
Both the dry-run and the real assembly were driven by a program which was also used to record the components of the panel in a dedicated database\footnote{For the chambers that will be installed in ATLAS the content of this database will be available to the reconstruction program.}.

\par The glue was distributed on the PCBs in less than 10 minutes by a Programmable Glue Dispenser hosted in the clean room, as shown in Fig.\ref{drift_panel_5}.
About 100 mL of glue are deposited on the five PCBs of each side of the panel, resulting in two layers of about 70 \mum\ thickness.

\myfig{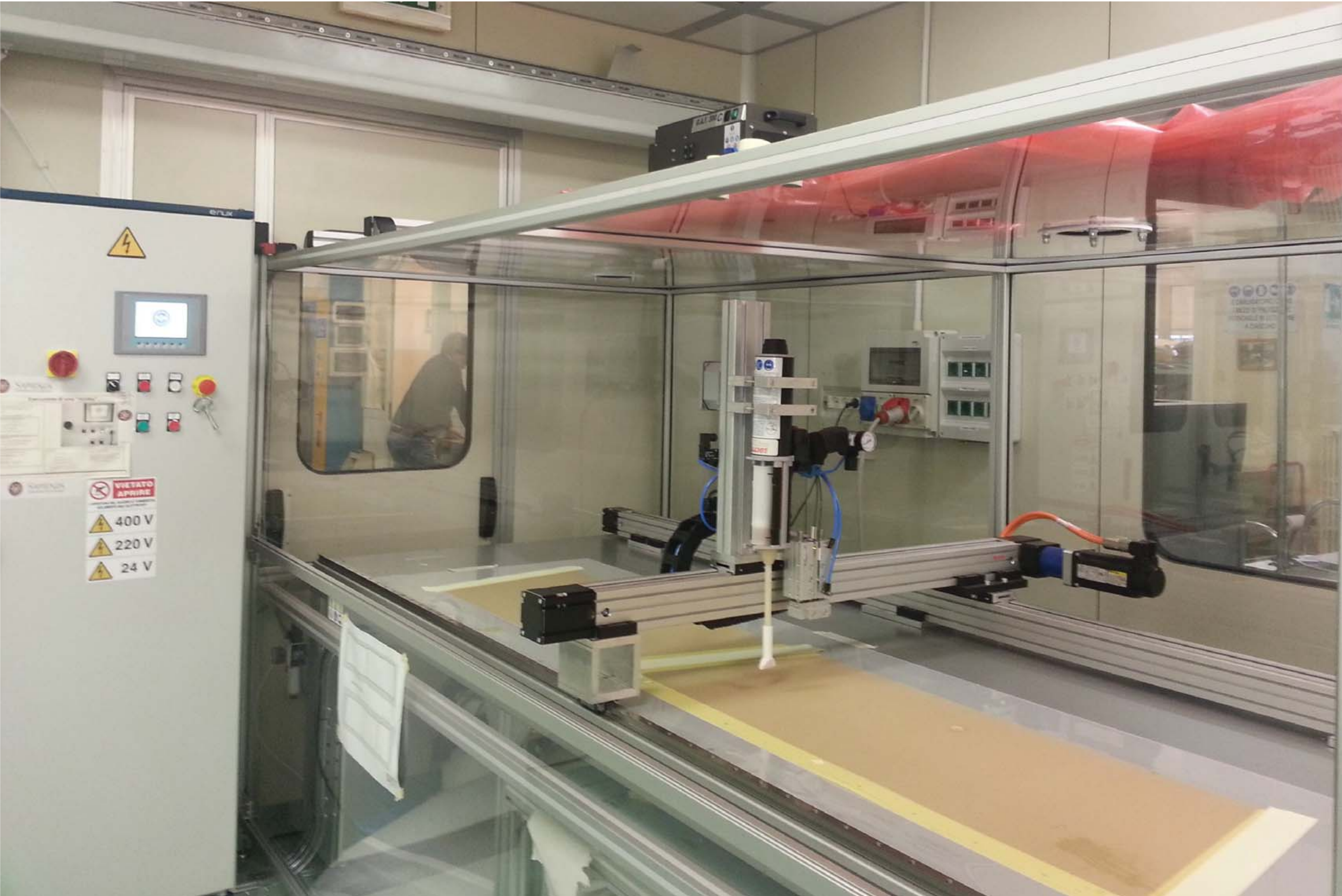}{76}{0}{The programmable glue dispenser}

\par The assembly of the panel was driven by an alignment bar mounted on the granite table.
First the glue was distributed on the PCB faces not covered by copper, which form the cathode plane of a MM layer.
These PCBs were then positioned on the granite table with the glue on the upper face and one of their lateral edges along the alignment bar.
The PCBs were fixed to the bar using 5 mm dowel pins.

\par Afterwards the glue was distributed on the lateral frames and the inner bars, which were then positioned  on the PCBs standing on the granite table.
The position of the lateral frames was determined with respect to the PCBs by 6 mm dowel pins.

\par At this point the five honeycomb pieces were positioned on the PCBs standing on the granite table.
The lateral frames, the inner bars and the honeycombs were all electrically connected together with small electric cables.

\par The following operations are different in the single- and two-step methods. In the single-step procedure, used for the outer drift panels, the glue was distributed on the second set of five PCBs of the panel, which were then positioned on the components already assembled on the granite table with the glue on the lower face.
Then an \myal\ mask was superimposed on the assembled panel.
Teflon dowel pins inserted in the holes in the mask determined the position of this second set of PCBs with respect to the other parts of the panel.
Then a cover was fixed by a special double face tape to the granite table, in order to form a bag with the table, as shown in Fig.\ref{drift_panel_1}.
An under pressure of 100-150 mbar was produced and maintained constant for 20 hours in the bag by means of a pump system, while the glue was being cured.
A computer assisted monitoring system recorded the pressure value inside the bag.
After one day, the cover was removed and the assembly of the panel was completed.

\par A more accurate procedure was followed to prepare the central drift panel, where a good planarity is demanded on both sides.
In this two-step procedure the second set of five PCBs was positioned without glue and the vacuum bag procedure was started.
When the curing of the glue on the bottom side was completed, the second set of PCBs was removed and the partially assembled panel was rotated and deposited on a second table.
The glue was dispensed on five PCBs, which were positioned on the granite table using the reference bar.
Afterwards the already semi-assembled panel was rotated and superimposed.
Finally, the vacuum bag was applied again for about 20 hours to complete the assembly of the central panel.

\subsection{Drift panel QA/QC} 
\label{drift-tests}

\par To measure the thickness and the planarity of the drift panels we have developed a small tool, called \emph{limbo} (Fig.\ref{drift_panel_6}), which consists of a rigid \myal\ profile, instrumented with 10 height gauges\footnote{Digimatic indicator Mitutoyo 543-790B\myreg.}, read out serially by a PC.
The gauges are zeroed on the granite table at the beginning of the measurement and then moved at pre-defined positions parallel to the bases.
The data is recorded automatically and analyzed both during the measurement and at the end.

\myfig{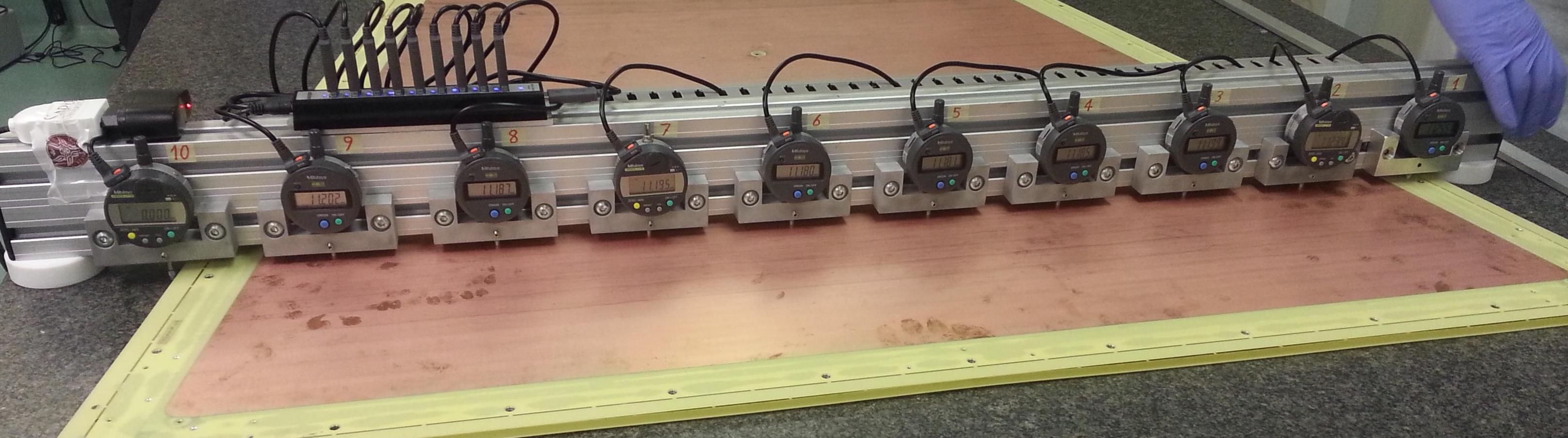}{76}{0}{The \emph{limbo} tool, while used to measure the planarity of a drift panel}

\par The drift panels of the \mymod\ have been measured twice, both with the \emph{limbo} tool and with a standard laser tracker at the LNF laboratory.
The results of two of these measurements are shown in Fig.\ref{drift_panel_7} and \ref{drift_panel_8a}.
The planarity is consistent with the required values, while the average thickness, which is smaller than the specifications, does not affect the chamber performances.
For the series production a thicker honeycomb will be used.
In addition, the measurements qualify the use of the Limbo tool for the QA/QC of the production.

\myfig{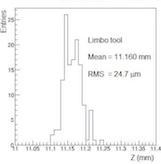}{76}{0}{Measurement of the thickness and planarity of the a drift panel with the \emph{limbo} tool}

\myfig{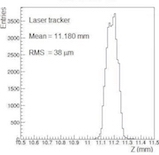}{76}{0}{The same plot as in Fig.\ref{drift_panel_7} with the laser tracker at LNF}

\par Electrical measurements were also performed on the assembled drift panels.
When applying 1 kV voltage the current measured was checked to be less than 10 nA.

\section{Drift Panels completion and Mesh assembly} 
\label{drift-comple}

\par After the construction and certification of the bare drift panels, they were transported to the \emph{Labo\-ratori INFN di Frascati (LNF)}.
The meshes (see Sec.\ref{layout}) were independently prepared and tensioned in the laboratories of the \emph{Sezione INFN di Roma Tre (RM3)}.
The stretched meshes, on their transfer frames, were also sent to LNF for the drift panel completion.

\par At LNF the drift panels were finalized, by finishing the mechanical construction, fastening the meshes and performing the final certification.
The positioning of the mesh at the right distance from the anode is guaranteed by the stretch of the mesh along the edges and by the electrostatic force due to the large electric field between mesh and anode that presses the mesh on the pillars.
This method, called \emph{floating mesh} \cite{floatmesh}, is a novel technique, since all previous MMs of smaller dimensions had been built with the so called \emph{bulk} technology \cite{bulkmm}.
In the floating mesh concept, the quadruplet can be re-opened since the mesh is not glued to the read-out PCB.

\mydummy{After the construction and certification of the bare drift panels, these panels were transported to the INFN laboratories of LNF, where also the mesh was following operations were performed:
\begin{enumerate}
\item Mechanical finishing:
\begin{itemize}
\item removal of the glue in excess, also check and reopening of the holes, when obstructed by glue;
\item fixing possible delamination problems;
\item gluing of inlet/outlet gas connectors\footnote{The design for the gas distribution inside the gap has been changed for the series production of SM1 modules: the gas is inserted from the major base of the panel and exhausted from the opposite base by means of a stainless steel pipes glued to the gas insert coming out from the drift panel.};
\item gluing of the interconnection drift spacers;
\item sealing of the PCB joints, and local gas-leakage tests (this operation was also applied to \myropans);
\item high voltage and electrical connections;
\item gluing of the mesh frame.
\end{itemize}
\item Mesh preparation and fastening:
\begin{itemize}
\item mesh stretching and punching for the interconnection holes (asynchronous with respect to the other operations);
\item washing/cleaning of the mesh;
\item gluing on the panel mesh frame.
\end{itemize}
\item Drift panels final certification:
\begin{itemize}
\item drift panels global gas tightness certification tests;
\item high Voltage tests;
\item mesh tension measurements.
\end{itemize}
\end{enumerate}
}

\subsection{Mechanical finishing of the drift panels} 

\par The finalization started with some minor fixing of the panels, like removing the glue in excess, checking and reopening the holes when obstructed by glue, fixing possible delamination problems, gluing of inlet/outlet gas connectors.

\myfig{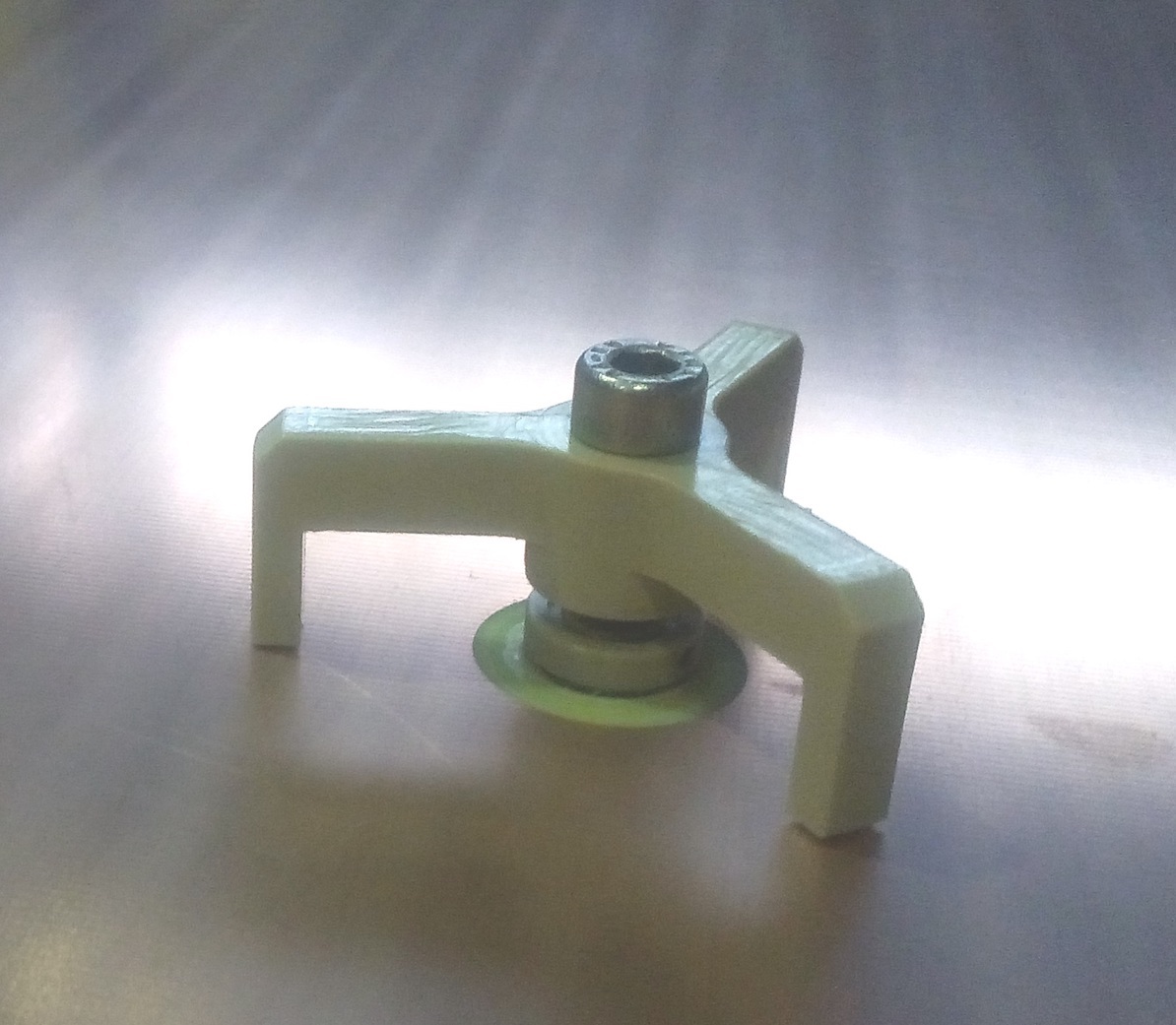}{78}{0}{The plastic tripod used to precisely position the interconnection drift spacers}

\par Then the interconnection drift spacers were glued with $\pm 25~\mum$\ precision in height.
To achieve this goal a special tripod was built.
The interconnection spacer was first fixed with screws to the tripod and then onto the panel, where the internal interconnection disk was placed.
The tripod was screwed onto the panel up to the point where the three feet touched its surface.
When the spacer was in place, a small amount of glue was distributed around the screw.
Since the feet have to slide on the copper surface and the tripod does not have to warp when tightened into the disc, the tripod was constructed of peek plastic\footnote{In the series production, the interconnection design has been modified, gluing the holder of the interconnection end cap inside the outer panel and replacing the plastic tripod by a stiffer one.}.
\mydummy{footnote: In the series production the interconnection design has been modified: in the external drift panels the holder of the interconnection end cap is glued inside the outer panel from the external face and soon after the interconnection spacer is screwed on it to the opposite side and then glued. The precision positioning scheme of the interconnection spacer is also changed. The plastic tripod has been replaced by a stiffer one. It is shown in Fig.\ref{mesh_00}: a precision stainless steel washer is placed concentric to the interconnection position; the interconnection spacer is screwed into the brass holder by the help of a dynamometric screw driver. Then this spacer is screwed to the panel by help of the dynamometric screw driver and then glued.}
Fig.\ref{mesh_00} shows the tripod, the interconnection spacer and the final assembly.

\myfig{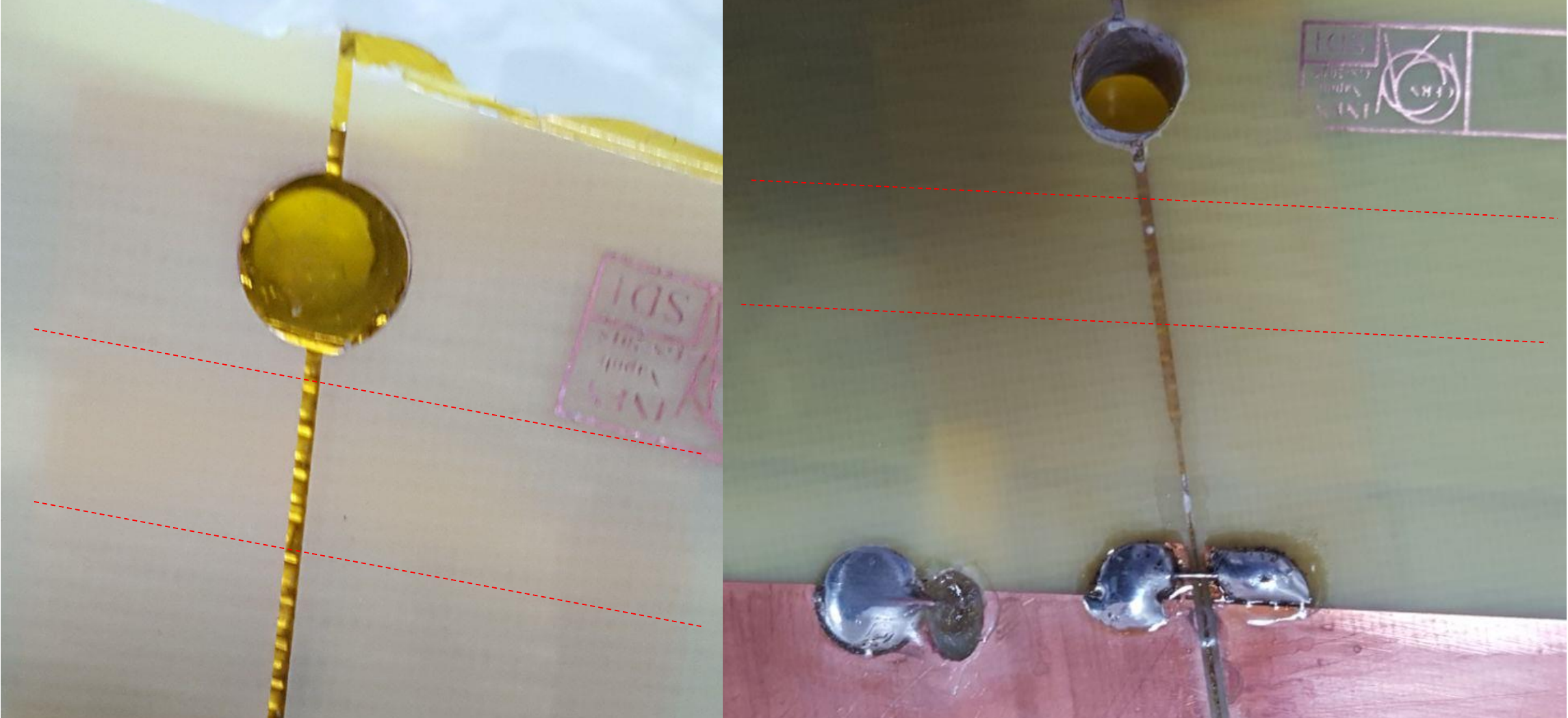}{78}{0} {A junction between two PCBs in the o-ring region (red dashed lines) before (left) and after filling with Araldite 2011 (right)}

\par All the junctions between two PCBs (four junctions for each side of a drift panel) were sealed where the o-ring was to sit, as shown in Fig.\ref{mesh_00b}.
This seal was performed by distributing a small thread of glue to completely cover that zone.
After few minutes the glue was heated with a heater and left to act for 10 minutes.
Then, with the help of a spatula, the glue was pushed inside the fissure, then the surface around the joint was cleaned with alcohol to remove glue residuals and covered with kapton tape.
The panel was finally turned upside down and left for 24 hours to cure\footnote{In the series production, this sealing is performed along the entire length of the joint.}.
\mydummy{longer footnote: add Adhesive paper tape is stretched 0.5 mm from the edge of the joint on both PCBs. The purpose of the tape is to calibrate the right amount of glue on the joint. The glue is prepared under vacuum and distributed along the entire length of the joint. Then it is left to act for half an hour and then with the help of a spatula pushed inside the fissure. At the end of this last operation the two strips of adhesive tape are removed and the panel left un-touched for 24 hours for curing.}

\par The mesh frames were then glued on the drift plane with the required accuracy of $\pm$200 \mum\ in the plane and $\pm$0.025 \mum\ in height, guaranteed both by the precision of the mesh frame profile and by plastic rulers referring to the closing holes.
To avoid an extra thickness due to the glue, the base of the mesh frame profile had a groove of 400 \mum\ depth, where the glue had to be precisely distributed.
The distributor filled the groove with 300 \mum\ depth glue.
After glue filling the profile is turned upside down and positioned on the panel by the help of the plastic positioner.
At this point the grounding screws were inserted.
To force the mesh frame profiles to adhere to the panel a heavy frame with closed-cell expanded elastomer as contact surface was placed on top of the frame itself during the 24 hours curing.

\subsection{The fastening of the mesh} 

\subsubsection{Mesh stretching} 

\myfig{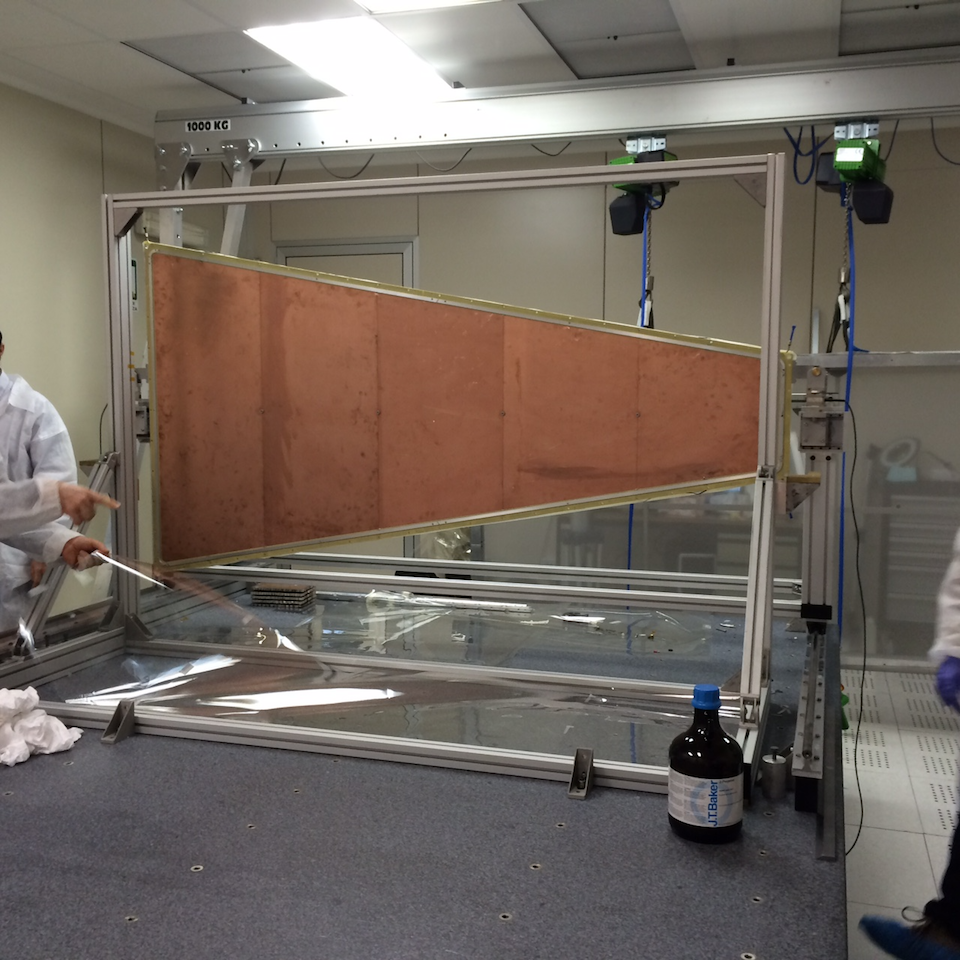}{78}{5}{Mechanically finished drift panel before gluing the mesh, hanging in vertical position in preparation for planarity measurements. The mesh frame fixed on the panel is visible all around the perimeter}

\par The electrical transparency of the stainless steel micro-mesh, necessary for the passage of drift electrons, depends by both its mechanical structure and the ratio between the amplification and drift electric fields.
To avoid amplification gap inhomogeneity, due to sags between the pillars or to wrong positioning, the mesh must be precisely tensioned and glued.
\par The mesh used for the Micromegas prototypes described in this paper had 28 $\mu$m wire diameter with 50 $\mu$m opening (denoted as 28/50), whilst the mesh used in the final version is 30/70.
\par During the construction, it was stretched to the desired tension, then prepared with holes where the interconnections must pass through, and finally glued on the \myal\ frames (the mesh frames) which were in turn fixed (glued and screwed) on the drift panel (Fig.\ref{mesh_01}).

\par A nominal mesh tension in the range $7-10$ N/cm with a uniformity of $\pm 10\%$\ is required in the drift panel after gluing.
Since, during mesh gluing, its tension increases by about 15\% due to it being pressed onto the frame, the initial tension on the transfer frame was adjusted to 8 N/cm.

\mybfig{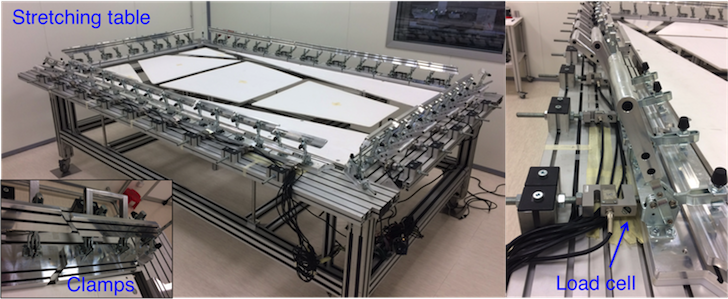}{130}{5}{Left: The mesh stretching table. In the inset: the exploded view of one of the clamps in open position. Right: The load cells monitoring the clamp pulling force}

\par A stretching table ($\sim 2 \times 3~\msq$) was built in the RM3 laboratory (Fig.\ref{mesh_02}).
The table is equipped with a total of 28 clamps, each 330 mm long, placed along the four sides. Short sides have 5 clamps while long ones have 9 clamps. Only half of the total clamps are equipped with load cells and can be pulled, the remaining clamps are fixed on the table. Moving clamps can be independently pulled through screwing nuts. Load cells are employed to measure the applied pulling force, which is acquired by an Arduino based DAQ system then monitored live on a display and stored on a computer.
The table also features a platform within the mesh area (white in the figure), in order to support the mesh before closing the clamps, and prevent the mesh falling down inside the table and any sagging during stretching.
It can be lowered after clamp closure in order to allow tension measurements of the mesh.
The tension was measured with a digital gauge\footnote{Sefar Tensocheck\myreg 100.}.

\myfig{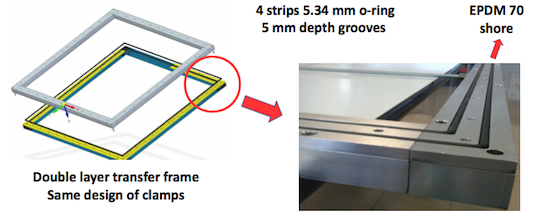}{78}{5}{The mesh transfer frame.
Left: the concept: the mesh is pressed between two frames closed with screws.
Right: one of the two layers with o-rings to maintain the mesh in position avoiding slipping.
The holes for the screws are visible between the o-rings}

\myfig{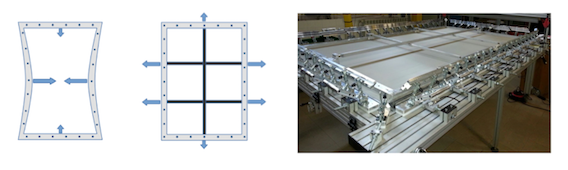}{78}{5}{Deformation of the mesh frame when the clamps are released.
Restoration of the mesh tension with adjustable pulling bars}

\par In order to move the mesh, while keeping it at its nominal tension, a set of reusable transfer frames were built.
As can be seen in Fig.\ref{mesh_03}, the mesh holding system is composed by two frames, both of them with double round of o-rings to clamp and maintain the mesh position.
The mesh was placed between the two frames on the stretching table, while the frames are blocked with screws.
The holes in the mesh between the o-ring, produced by the tightening screws, have no effects on the tension.
When the mesh was well-tightened in the frames, the clamps were released.
Since, as sketched in Fig.\ref{mesh_04}, the mesh tension can bend the transfer frame, three adjustable pulling bars are used to restore the shape of the frame and therefore the correct mesh tension.

\myfig{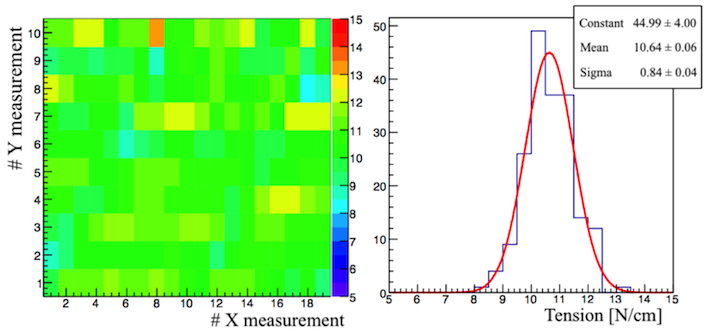}{78}{5}{Mesh tension measurement after stretching, on the transfer frame}

\par Once the mesh was ready, stretched and in position in the transfer frame, a full map of the tension was produced.
Fig.\ref{mesh_05}\ shows the tension measurement of one of the meshes after the transfer process, following the mentioned procedure.
A uniformity of about 7\%, sufficient for the detector requirements, was reached.

\subsubsection{Mesh punching and gluing on the drift panels} 


\par In order for the interconnections to pass through all the panels in the assembled quadruplets, the meshes were perforated, using a punch through tool after local passivation.
This procedure, which prevents mesh fraying and/or filaments to stick-out was produced with a little ``button'' of glue, typically 10 mm in diameter and 150 \mum\ in thickness.
The glue was first dispensed on a mylar foil, and then pressed and left curing on the mesh.

The average mesh thickness on the passivated region was around 65 \mum, about 10 \mum\ in excess with respect to the bare mesh.
The mesh punching (Fig.\ref{mesh_07}) was done after the passivation glue curing (24 hours).

\myfig{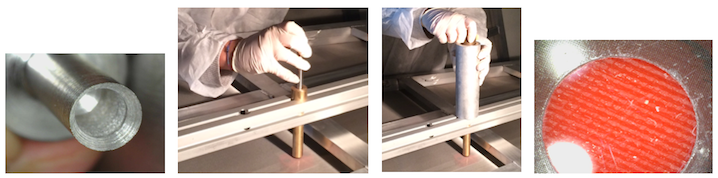}{78}{5}{Mesh punching: From left to right: (1) the hollow punch tool; (2) the tool inserted in a brass guide; (3) a weight dropped from a calibrated height on the hollow punch; (4) the perforated mesh}

\par The final step in the drift panel completion was the transfer and gluing of the pre-stretched mesh from the transfer frame onto the panel mesh frame.
The glue was deposited on the inclined surface of the mesh frame with the help of a dedicated tool.
Particular care was taken during the gluing procedure to avoid any glue on the top part of the mesh frame, in order to keep the mesh at the proper height with precision, and also to avoid the glue from sticking out from the inclined surface.
The glue filled the grooves, with an extra-height of about 80 \mum.



\mybfig{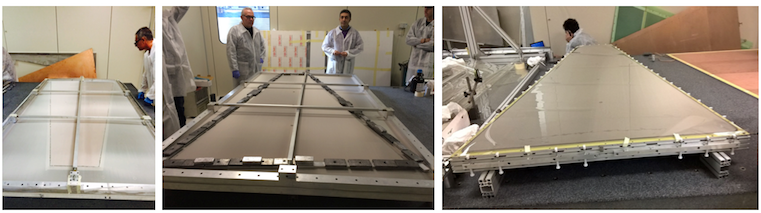}{160}{5}{Left: The fastening of the mesh onto the drift panel. Centre: Curing of the glue attaching the mesh onto the drift panel. Right: Mesh glued on the drift panel}

\par Once the glue was distributed all along the mesh frame, the stretched mesh on its transfer frame was lowered onto the drift panel (Fig.\ref{mesh_09}\ left).
This operation was done placing a 5 kg lead bricks on the mesh, all around the perimeter of the panel (Fig.\ref{mesh_09}\ centre).
After curing overnight, the mesh was cut with a sharp scalpel, and the final product was ready for the final certification measurements and for quadruplet assembly (Fig.\ref{mesh_09}\ right).

\subsection{Final certification} 

\subsubsection{Mesh tension measurements} 

\par The mesh tension was measured on the four meshes stretched and glued on the two external drift panels and on the two sides of the central panel.
All measurement results were within requirements, or only marginally outside\footnote{For the series production, procedure and tool improvement allow for results well within the requirements.}.
In Fig.\ref{mesh_10}\ the measurements corresponding to the external drift panel 2 and to the two sides of the central panel are reported\mydummy{The data of the first external panels went lost.}.

\myfig{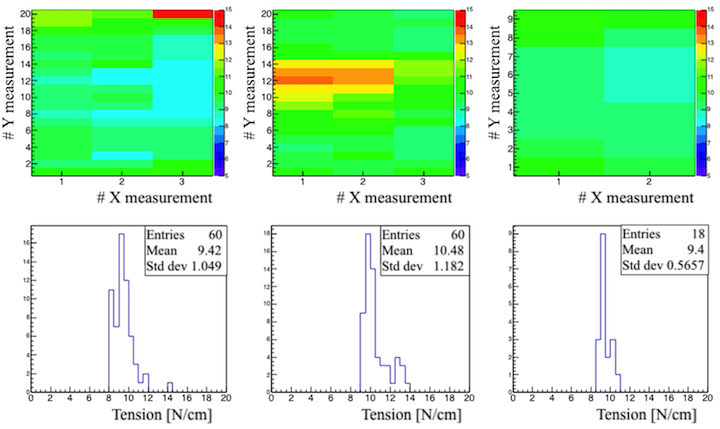}{78}{5}{Mesh tension maps (top row) for three out of four stretched meshes and corresponding distributions (bottom row).
On the left column, the map and distributions of the external drift panel 2 is reported.
The maps and distributions for the 1 and 2 sides of the central drift panel are reported in the central and right column respectively}

\subsubsection{Global gas tightness certification tests} 

\par After the drift panel was sealed, the local gas leakage checks performed with success and the mesh-frame and the gas distribution pipes mounted and glued, the tightness certification of the entire panel was performed.
For this test, a couple of gas-tight \myal\ dummy panels were used.
They served as a vessel for the drift panel and measured the leakage rate after filling the gas-gap with air.
\mydummy{The central drift panel was coupled with both dummy panels, while the outer one was coupled with only one of them.
The o-ring was then mounted onto the drift panel where the gas-gap frame is screwed in the final position to form the o-ring groove.}

\par The sequence for coupling an outer drift panel with a dummy panel was the following:
\begin{enumerate}[a)]
\item the outer panel, equipped with mesh and gas gap frames, was placed on the gas tight table;
\item the o-ring was positioned on the groove of the panel;
\item a dummy panel was positioned in such a way to close the drift panel, creating the gas gap;
\item the outer drift and dummy panels were clamped together;
\item a structure made of aluminum bars, parallel to the panel bases, was tightened on top of the dummy panel, in order to reduce the deformation of this panel when the gas gap is filled with an overpressure of air of few mbar.
\end{enumerate}
\par \noindent Then, about 100 mL of air were inserted by means of a syringe into the gas gap and the leak rate deduced by the pressure decay rate method.
The differential pressure of the gas-gap was monitored for at least two hours.
If the air leak rate exceeded the maximum allowed limit, the cause of it was carefully investigated and repaired and the gas tightness test was repeated.
The resulting leak rate of a panel was typically 0.10 mL/min, quite consistent with the requirements.

\section{Module Assembly} 
\label{assembly}

\par The finalized drift and \myropans\ were then vertically assembled in a class 10,000 clean room at LNF, equipped with a certified granite table with $11~\mum$ measured planarity.
The mechanical tools used for the procedure described in this section are shown in Fig.\ref{assembly_01}.

\mybfig{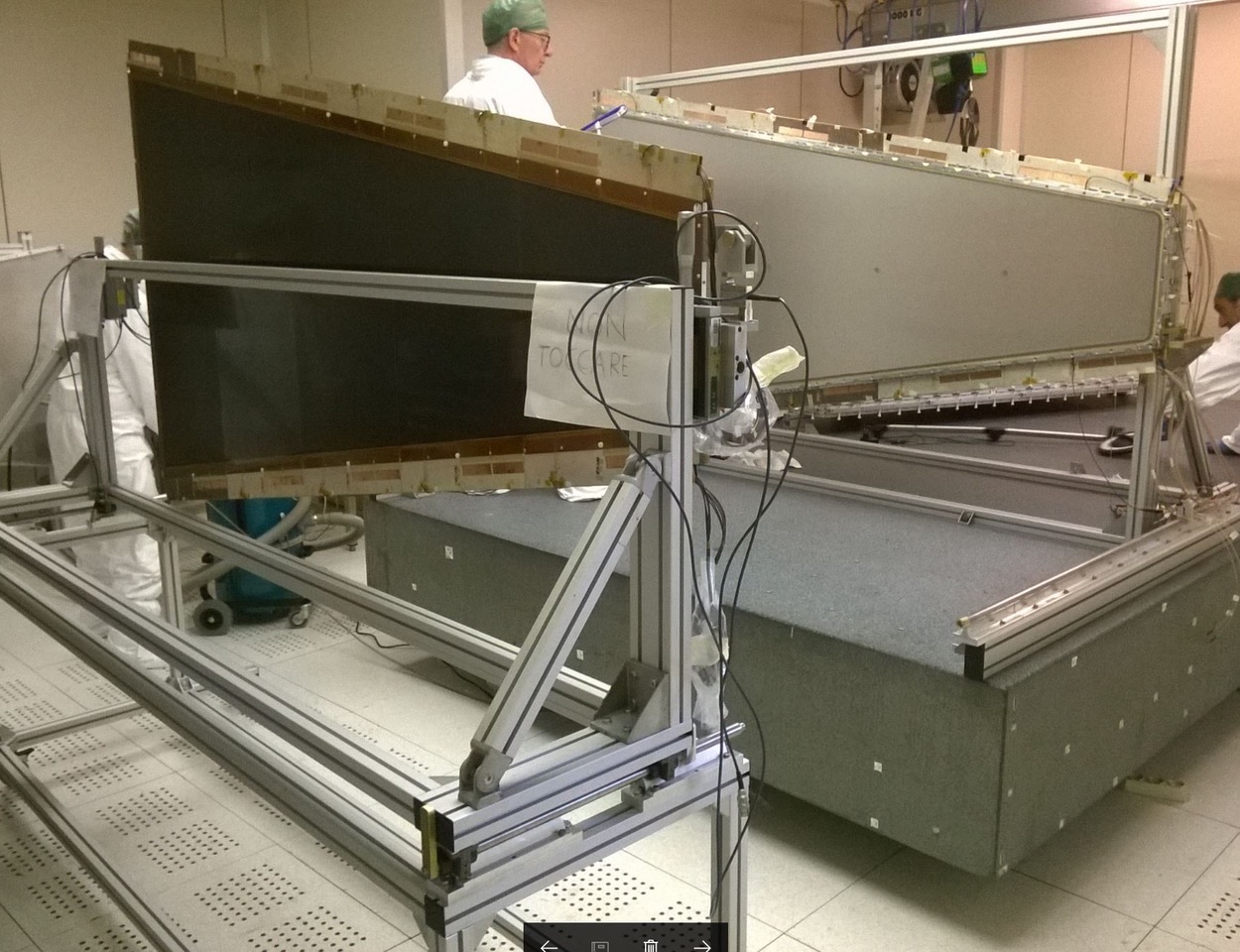}{140}{0}{The granite table used for assembly, equipped with the assembly tool. The photo shows the drift panel 5 mounted on the table and the \myropan\ 4 ready for mounting}

\par The five panels presented in Fig.\ref{layout_3} were mounted and assembled, starting from the external drift panel 5, up to the external drift panel 1, as described in the following.

\subsection{Drift panel 5}

\par The external drift panel 5 was positioned on the \emph{stiff-frame}, a mechanical structure used to guarantee the panel planarity during the assembly procedure.
The stiff-frame was produced using a commercial \myal\ profile glued with \myal\ brackets with mechanical tolerances of $\sim 100$ \mum.
The panels were accurately positioned on the frame and fastened with plastic brackets, as shown in Fig.\ref{assembly_02}.
The alignment of the panel on the stiff-frame was obtained by adjustment screws, shown in Fig.\ref{assembly_03}.
Then the stiff-frame with the panel was fixed on the assembly tool.

\myfig{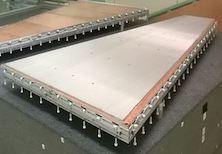}{50}{5}{Two panels positioned on their stiff-frames with plastic brackets}

\myfig{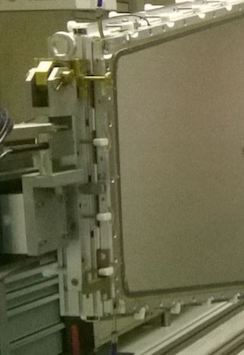}{50}{5}{One of the adjustment screws for the panel alignment}

\subsection{\myropan\ 4} \label{ropanel}

\par The stereo panel 4 was added by mounting brackets to hold it on both sides (as shown in Fig.\ref{assembly_04}).
The bracket on the long side was equipped with spherical joints, while on the short side there was a simple support piece.
The panel was then positioned on the assembly tool as shown in Fig.\ref{assembly_01}.
Once the drift and the \myropan\ were positioned face to face, the two panels were subject to a cleaning procedure, in order to remove dust from the surface of the PCBs and the mesh with a dedicated electrostatic roller, while using an ionized nitrogen gun.
After cleaning, the HV test was performed\footnote{All the HV test during the assembly were performed in air by applying 750 V and requiring a current smaller than a few nA.} on the \myropan\ surface with a mesh dedicated tool.
If critical regions, shorts or poorly insulated regions were found, they were protected with kapton tape.
Ten pins with 2 mm diameter were then inserted for each layer to align the Front-End boards.
Finally, the alignment between the stereo panel and the drift panel was checked using two delrin\myreg ~pins (6 mm diameter), inserted in the corresponding holes.
The two panels were then connected using expansion rods, as shown in Fig.\ref{assembly_05}.
The expansion rods were designed to fix the panels and to ensure the o-ring compression.
Finally, another HV test and a gas leakage test were performed.
A section drawing of the assembled panels is shown in Fig.\ref{assembly_06}.

\myfig{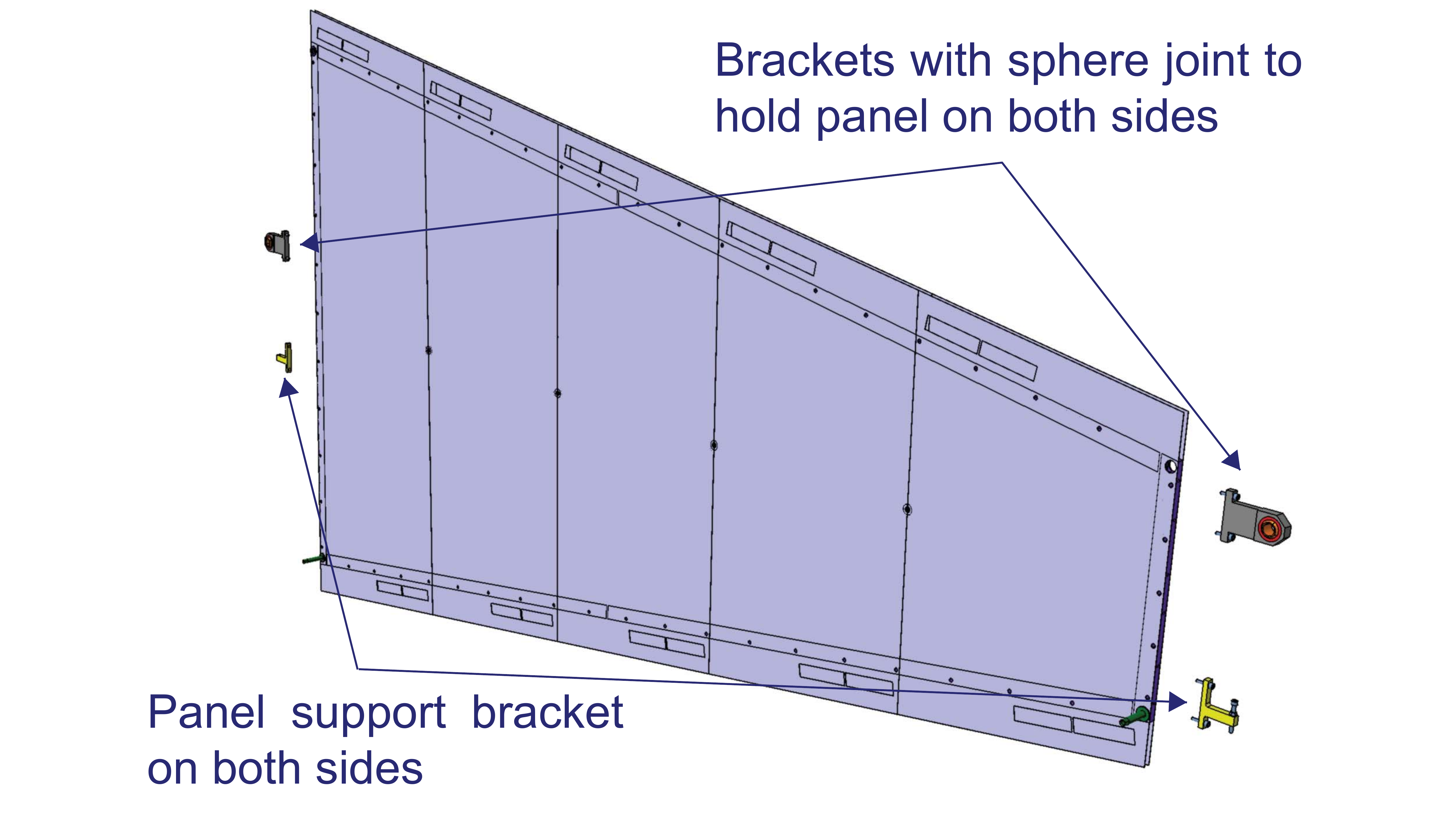}{77}{5}{The mounting of a \myropan}

\myfig{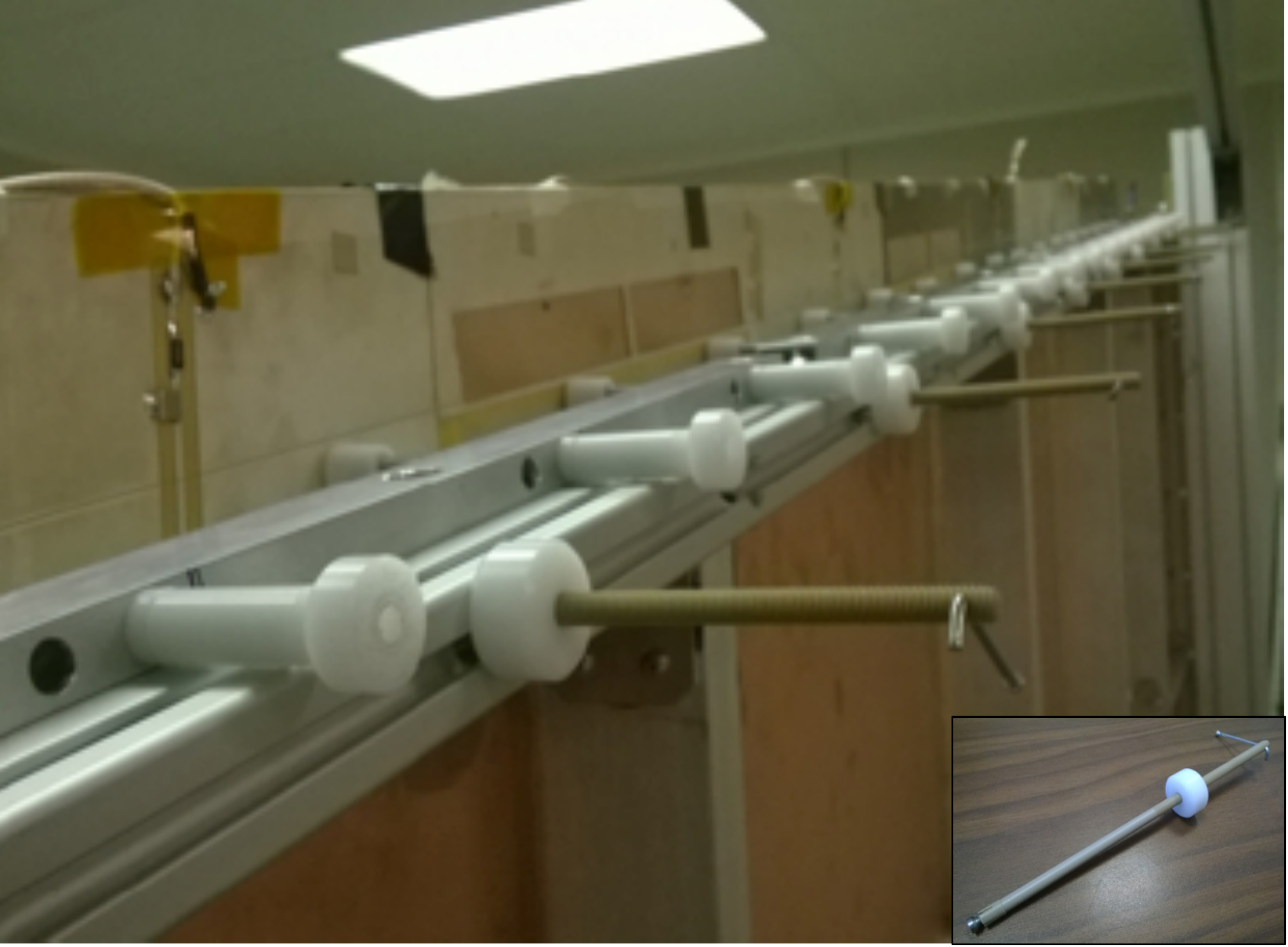}{77}{5}{The expansion rods}

\subsection{Drift panel 3}

\par 
The central drift panel 3 was added at this stage. Brackets with spherical joints and support brackets were mounted to hold the panel on both sides, then the panel was positioned on the assembly tool, facing the stereo panel 4.
After cleaning, the panel was aligned with respect to the stereo panel screw holes, checked with at least two delrin\myreg ~pins (6 mm diameter), and then fixed to the assembled structure using expansion rods on the second half screw holes.
HV and gas leakage tests were then performed.

\myfig{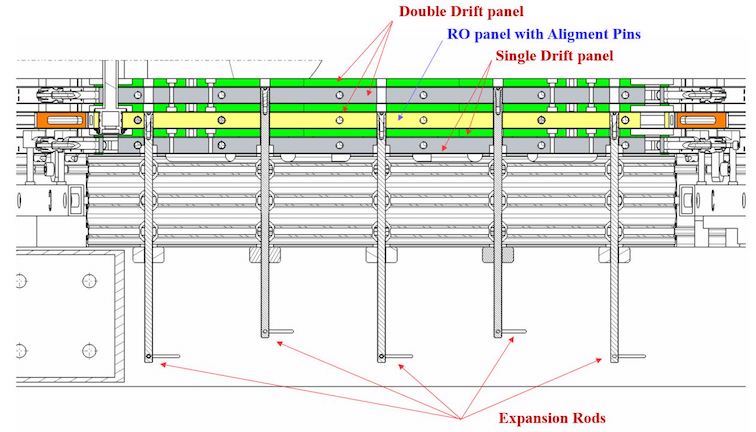}{77}{5}{A section drawing of the assembled panels 5, 4 and 3}

\subsection{\myropan\ 2}

\par The assembly of the eta panel 2 began by equipping the panel with brackets, as explained in Sec.\ref{ropanel}.
The panel was then positioned on the assembly tools and subject to the described cleaning, HV and gas tests.
The alignment procedure, shown in Fig.\ref{assembly_07}, consists of the alignment of the sides of panel 2 with respect to the panel 4 by means of alignment pins.
The usage of load cells on both sides of the panel 2 allows to avoid forces on pins due to misalignment between the panels.
This procedure was achieved by monitoring the weight load 
and by controlling the panel position with micrometric screws.
Weights of $200-300$ g, corresponding to a pin deflection of about of 20 \mum , were considered acceptable.
Once these values were achieved on both sides, the panel was fixed and the HV and gas-leakage tests were performed.

\myfig{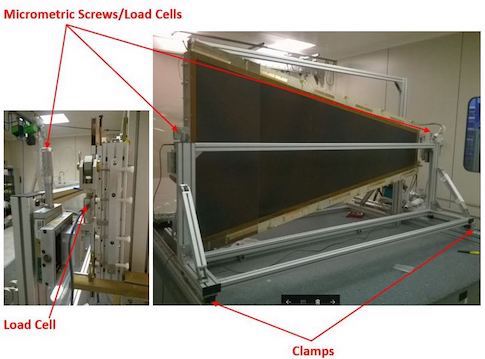}{77}{5}{The RO panel 2 alignment procedure, using micrometric screws and load cells}

\subsection{Drift panel 1}

\par Finally, the external drift panel 1 was assembled.
Similarly to panel 5, it was fixed on the stiff-frame and mounted on the assembly tool.
Then it was faced to the panel 2, cleaned and HV tested with the mesh tool.
Before the module completion and the final test, interconnection plugs with o-rings on both sides were inserted.
HV and gas-leakage test on the full module were performed.
All the brackets between the external drift panels and the stiff-frame were removed and the expansion rods were substituted, using a dynamometric key, with final screws.

\par The module was then dismounted from the assembly tool as shown in Fig.\ref{assembly_08b}.

\myfig{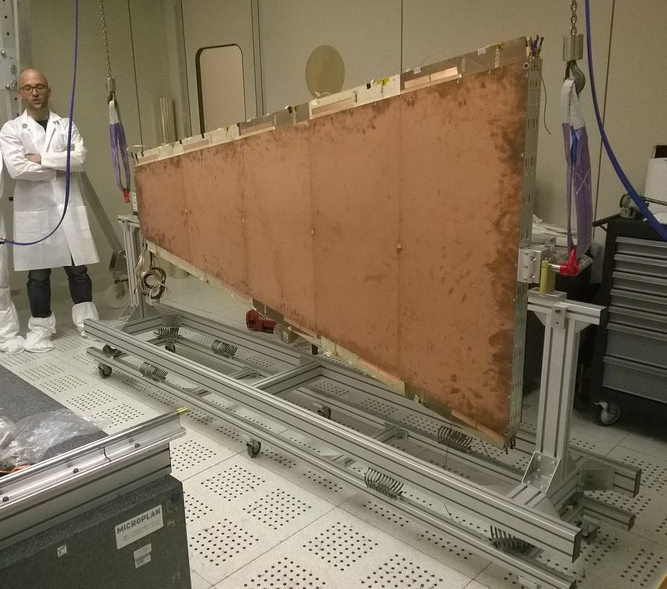}{77}{5}{Dismounting the assembled quadruplet}

\subsection{Quality tests}

\par After the completion of the assembly, a final HV test with the final gas mixture was performed on the complete chamber.
When a voltage of 590 V was applied, the current in some sectors of the chamber was exceeding the allowed values.
This area, corresponding to approximately 20\% of the chamber, was excluded from the subsequent measurements, as discussed in section Sec.\ref{testbeam}.

\section{Test-beam results} 
\label{testbeam}

The performances of the \mymod\ were then investigated with the CERN H8 particle beam.
The purpose of the test beam was to evaluate the performance of the chamber in terms of spatial resolution, strip alignment and efficiency. In the following the most relevant results are presented.

\subsection{Beam line and detectors} 

\myfig{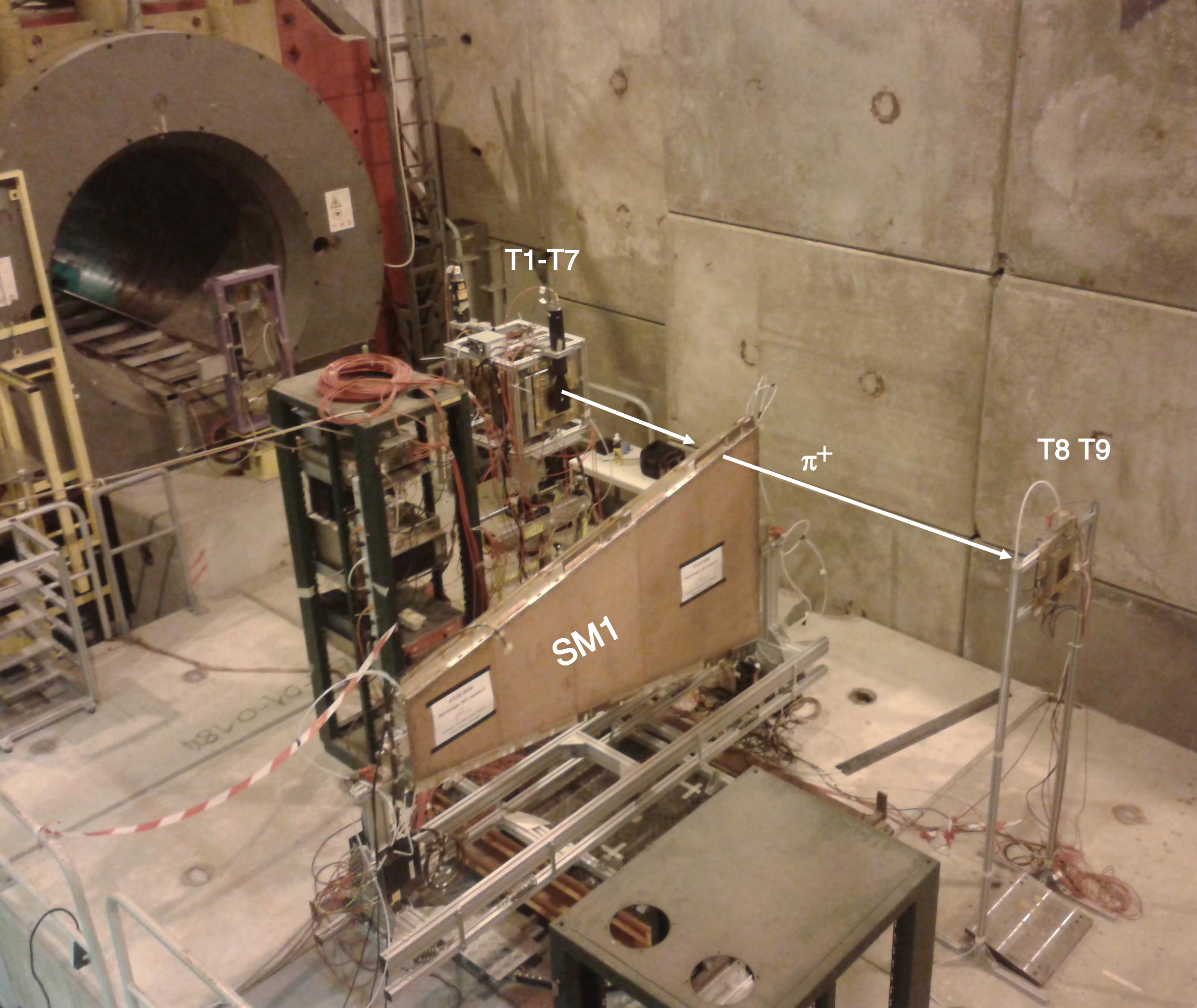}{76}{0}{The \mymod\ on the platform}

\par The measurements here described were performed at CERN in June 2016 at the H8 beam line of the Super Proton Synchrotron (SPS).
A 180 GeV/c $\pi^+$ beam with a rate ranging from 1 kHz to 500 kHz and with a beam spot of about 1$\times$1 \cmsq\ was used.

\par The chamber was filled with a Ar:CO$_2$ (93:7) gas mixture at atmospheric pressure with a flux rate of 20 L/h.
Data have been collected at different HV amplification values from 550 V to 580 V, while the HV drift was set at -300 V.
When not explicitly mentioned, the results presented here are given with a voltage applied on the resistive strips at 580, 560, 570, 580 V for layers 1 to 4 respectively.

\par To perform position scans the \mymod\ was placed on a movable platform, as shown in Fig.\ref{tbeam_1}.
The experimental setup was composed by an array of detectors of which the \mymod\ was part, as shown in Fig.\ref{tbeam_1} and Fig.\ref{tbeam_2}.
Nine small dimension bulk MM chambers (Tmm or separately T1-T9), some of which with two dimensional coordinate read-out, were used as a reference.
In the presented data the beam was centered either on PCB3 or on PCB5 of M0, in both cases performing a two-dimensional scan in order to test the chamber in different positions.
Only data with beam direction orthogonal to the chamber plane are considered in this paper.

\myfig{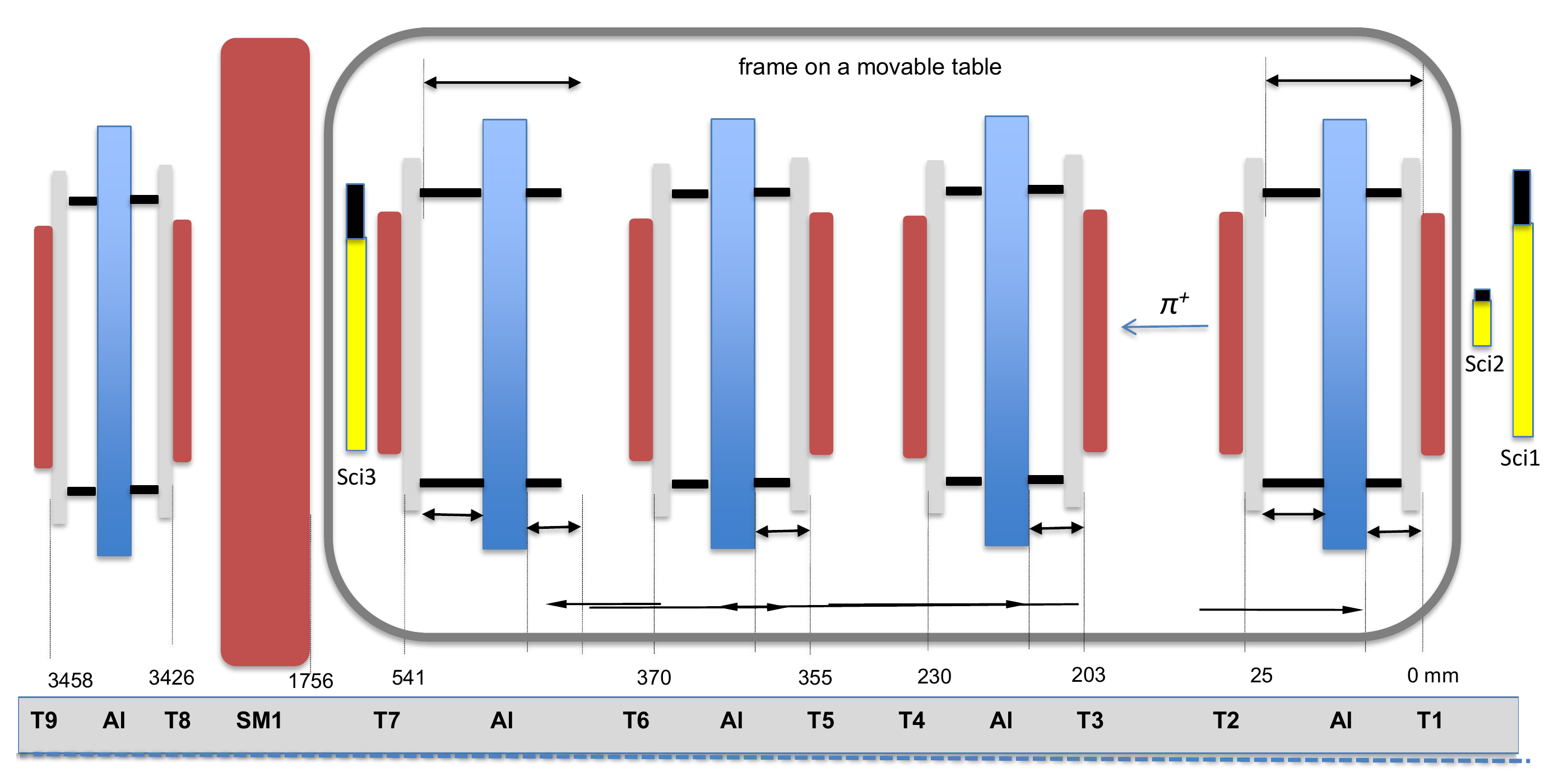}{76}{5}{Test beam setup (not to scale)}

\subsection{Output signal and data reconstruction} 

\par In the final NSW setup the MM read-out chain\cite{nswtdr} will be based on VMM ASICs. On the test beam the read-out was provided by APV25 and SRS (Scalable Readout System) modules operating at 40 MHz, the same read-out used for all previous tests done on smaller size MM chambers. 
The APV25 \cite{apv25} is a 128 channel chip, whose output data structure consists of the values of the charge collected in each sample, while the SRS is a module used to interface the APV25 signal output.
An APV25 channel was connected to each strip of the \myropans.
For each event and each channel, a total of 21 samples, one every 25 ns, was recorded in an acquisition time window of 525 ns. 
We stress here that the APV25-SRS chain is not optimized for the read-out of large chambers in a high rate environment.

\par The raw data have been processed by a chain of C++ codes, developed for the MM test beam.
Only events with a single well reconstructed track in the Tmm chambers have been analyzed.
For the remaining events, the values of the charges in the single strips of the \mymod\ have been pedestal subtracted and corrected to consider known cross-talk effects between the electronic channels.

\par The time of the signal has then been computed for each channel, by fitting the 21 charge samples as a function of time with a Fermi-Dirac distribution.
The half-height parameter of the function ($t_0$) is considered as the time of the signal in the strip and used in the reconstruction.
A group of nearby channels with compatible times, possibly hit by the same track (a \emph{cluster}) is identified.
The track coordinate is measured as the cluster centroid, computed by weighting the position of each channel with its charge\footnote{For tracks orthogonal to the chamber plane, we do not use the alternative \emph{\mTPC\ method}, which consists in fitting a straight line to nearby channels, using the $t_0$'s and a constant drift velocity\cite{nswtdr}.}.

\subsection{Spatial resolution} 

\par Measurements of the spatial resolution of the SM1 layers have been evaluated both for the precision coordinate (\myeta, for coordinate definition see Sec.\ref{layout}) and for the second coordinate (\myphi).

\par For the \myeta\ coordinate, the resolution is computed as the difference in position between the two eta layers of \mymod, divided by $\sqrt{2}$.
This method assumes that the planes have the same resolution and the beam divergence is negligible.
The data from PCB5, together with a gaussian fit, are shown in Fig.\ref{tbeam_3a}.
The resulting intrinsic resolution for the \myeta\ coordinate is found to be 81 \mum.
As a check of the result, Fig.\ref{tbeam_3b} presents the same plot for PCB3, showing an intrinsic resolution for the \myeta\ coordinate of 90 \mum.

\myfig{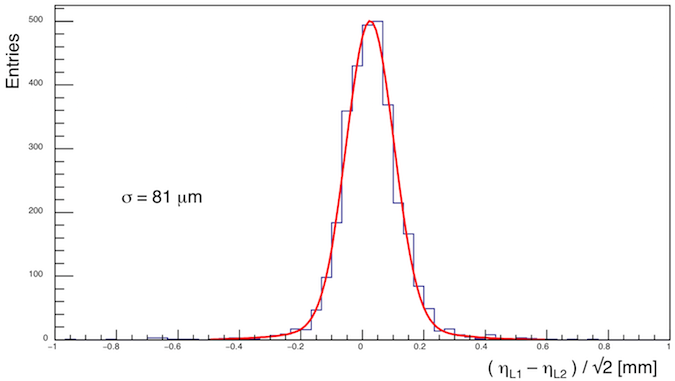}{76}{0}{Spatial resolution of the \myeta\ precision coordinate for PCB5}

\par For the \myphi\ coordinate, the resolution is computed as the signed difference between the position reconstructed by the stereo layers of \mymod\ and the position extrapolated from the Tmm reference chambers.
The residuals and the gaussian fit are shown in Fig.\ref{tbeam_4}.
The resulting intrinsic resolution for the \myphi\ coordinate has been found to be 2.3 mm.

\myfig{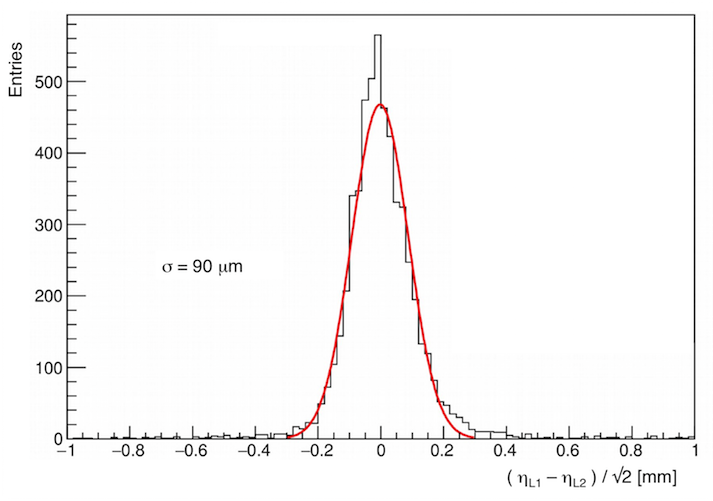}{76}{0}{Spatial resolution of the \myeta\ precision coordinate for PCB3}

\myfig{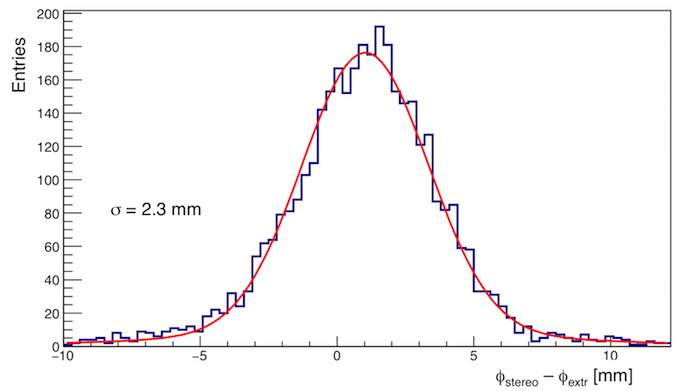}{76}{0}{Spatial resolution of the second coordinate \myphi\ for PCB5}


\subsection{Strip alignment} 

\par A possible misalignment of the read-out strips is an independent source of resolution error and is not controlled by the previous checks, which rely on local measurements.
Therefore, in order to have an indication of layer-to-layer rotation or strip pattern global deformation, a determination of the strip displacement in the \myeta\ coordinate as a function of the \myphi\ coordinate has been performed for the different layers of PCB5 and PCB3.
The measurement is performed by aligning the first layer with respect to tracks reconstructed in the reference chambers and then looking at the \myeta\ coordinate reconstructed from the layer 2 and the one reconstructed combining the information from the stereo layers.
The results obtained performing this measurement for PCB5 (PCB3) are shown in Fig.\ref{tbeam_5} (Fig.\ref{tbeam_6}).
The displacements reported for both PCB5 and PCB3 are in all cases within $\pm 70~\mum$ in the explored range. Moreover, in the case of PCB5 a clear slope is observed, indicating a slight rotation at the level of 0.1 mrad between the strips of L1 and the strips of the other layers.

\myfig{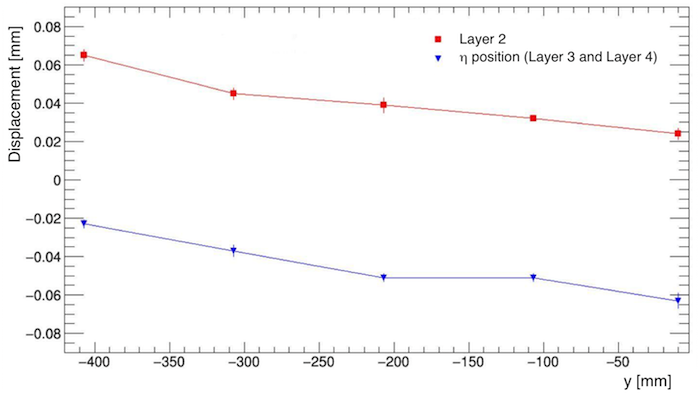}{76}{0}{Strip displacement as a function of the quoted Y coordinate on PCB5. In red, average residuals of precision coordinate reconstructed in layer 2 with respect to layer 1. In blue, average residuals of precision coordinate reconstructed in stereo layers with respect to layer 1}

\myfig{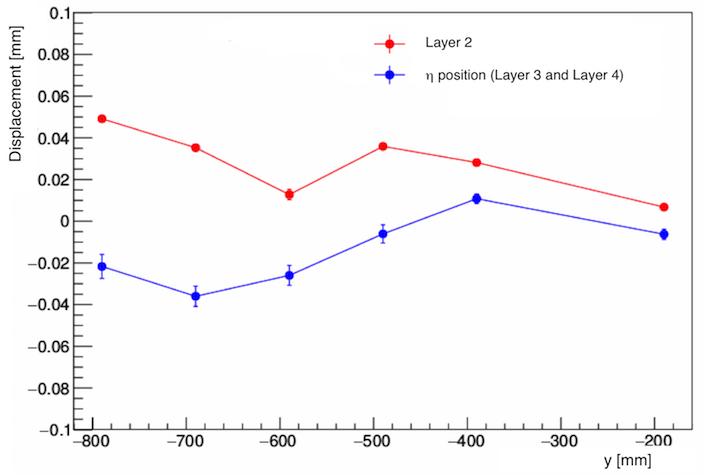}{76}{0}{Strip displacement as a function of the quoted Y coordinate on PCB3 (see caption of Fig.\ref{tbeam_5}) }

\subsection{Efficiency of \mymod\ Layers} 

The efficiency of the \mymod\ response has also been studied.
Starting from tracks reconstructed with the external Tmm chambers, for each layer of the \mymod\ we define a one-dimensional interval of $\pm$ 1.5 mm around the extrapolated point.
The efficiency is defined as the ratio ${N_{in}}/{N_{tot}}$, where $N_{in}$ is the number of tracks with at least one cluster in the interval and $N_{tot}$ is the total number of considered tracks.

\par In Fig.\ref{tbeam_7} the efficiency as a function of the track position is shown for the first layer.
As shown in the figure, the average value is $\sim98\%$.
The efficiency as a function of the HV amplification voltage for all the layers of PCB5
is shown in Fig.\ref{tbeam_8}.
From the figure we notice that an efficiency plateau with values between 93 and $98\%$ is reached for all layers at HV above 570 V. The small drop in efficiency observed at 580 V for layers 2 and 3 is due to HV instabilities observed for the two layers, that can actually be safely operated only up to 570 V. With respect to bulk small size MM prototypes the operating voltage is larger by about 50 V. This is interpreted to be due to an higher effective thickness of the amplification gap due to the floating mesh concept.

\myfig{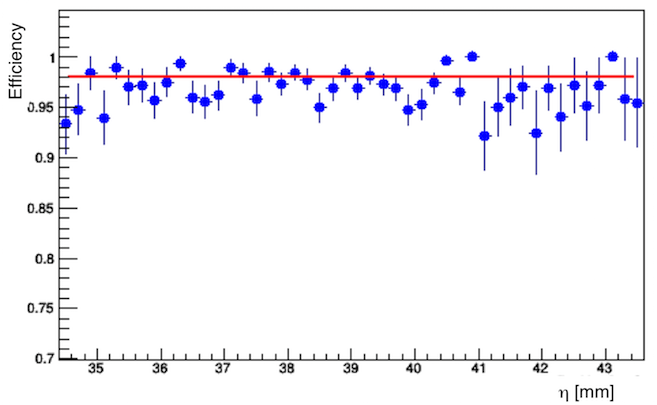}{76}{0}{Efficiency as a function of the track position for layer 1 at 580 V}

\myfig{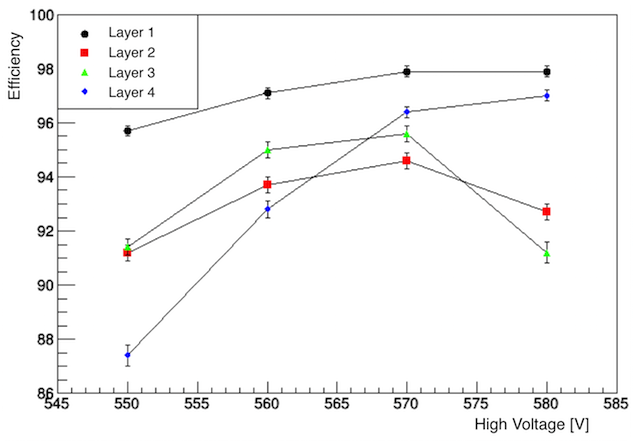}{76}{0}{Efficiency for all layers of PCB5 as a function of the HV}

\section{Conclusion} 
\label{conclu}

\par We have built a full size prototype of a Micromegas chamber for the ATLAS New Small Wheel and tested it with a beam. 

\par Although the components and assembly procedures are rather preliminary, the chamber performances are in
fair agreement with the tight requirements imposed by the LHC operations.
The mechanical assembly shows an accuracy in the strip position of the order of 100 \mum.
The planarity of the panels is measured to be about 40 \mum.
The chamber has been tested for global gas tightness and HV, showing that a large fraction of its area satisfies the required conditions.
The test beam operations have allowed for a measurement of the chamber performances.
Using a preliminary reconstruction algorithm, we have measured a resolution of about 80 \mum\ in the precision coordinate and 2 mm in the second one, with a single layer efficiency of about 95\%.

\section*{Acknowledgments}
\label{akno}
\par This work was performed within the ATLAS New Small Wheel Collaboration.
We are grateful to our ATLAS friends for helpful discussions and advice.
We thank the NSW management, in particular the project leader Stephanie Zimmermann for her continuous encouragement and guidance.
We would like to thank both the Technical Coordination of the NSW, led by Patrick Ponsot, and the Design Working Group for their effort to bring this project together.
We would also like to thank the CERN RD51 Collaboration and in particular Rui de Oliveira for their support in the chamber design.
In addition, we would like to thank the PCB production company, ELTOS and the PCB QA/QC group at CERN for ensuring the good quality of such a complex object.

\par The effort of assembling together the many components of the system was made possible by the dedication and skill of the technical staff of the mechanical and electronic workshops of our laboratories.
In particular we warmly thank the retired staff M.~Cappone, A.~Iaciofano, F.~Innocenti, M.~Petruccetti, C.~Piscitelli.
We are grateful to M.~Paris and the metrology service of the LNF and to L.~Tortora and the materials diagnostic service of RM3 for their impeccable services.

\par The test of the module was made possible by the efforts of the muon test beam group, which we warmly thank.
We also would like to thank the CERN-SPS team for its efficient and smooth operation of the beam line during our tests.

\section*{\refname}
%

%


\end{document}